\documentclass[sigconf,nonacm,screen]{acmart}
\IfFileExists{libertine.sty}{}{\usepackage{mathptmx}}
\usepackage{amsmath,bm}\usepackage{graphicx}
\usepackage{booktabs}
\usepackage{tikz}
\usetikzlibrary{arrows.meta,calc}
\usepackage{pgfplots}

\usepgfplotslibrary{groupplots}

\pgfplotsset{compat=1.18}
\definecolor{stretchOrange}{HTML}{BD7425}
\colorlet{flipc}{stretchOrange}    %
\definecolor{stretchRed}{HTML}{B44B45}
\colorlet{twistc}{stretchRed}      %
\colorlet{energyc}{blue!60!black}    %

\definecolor{branchGreen}{HTML}{177D68}
\definecolor{branchBlue}{HTML}{3569A8}
\definecolor{branchOrange}{HTML}{BD7425}
\definecolor{branchPurple}{HTML}{98517E}
\definecolor{figureInk}{HTML}{28313B}
\definecolor{figureMuted}{HTML}{697580}
\definecolor{boxEdge}{HTML}{9EAAB6}

\renewcommand\footnotetextcopyrightpermission[1]{}
\newcommand{\mbF}{\mathbf F}
\newcommand{\mbR}{\mathbf R}
\newcommand{\mbS}{\mathbf S}
\newcommand{\mbU}{\mathbf U}
\newcommand{\mbV}{\mathbf V}
\newcommand{\mbC}{\mathbf C}
\newcommand{\mbP}{\mathbf P}
\newcommand{\mbH}{\mathbf H}
\newcommand{\mbI}{\mathbf I}
\newcommand{\mbv}{\mathbf v}
\newcommand{\cof}{\operatorname{cof}}
\newcommand{\tr}{\operatorname{tr}}
\newcommand{\vct}{\operatorname{vec}}

\newcommand{\SO}{\mathrm{SO}(3)}
\newcommand{\secref}[1]{Section~\ref{sec:#1}}
\newcommand{\eqn}[1]{Equation~\eqref{eq:#1}}

\title{On the Singularities of Corotated Elasticity}
\author{Floyd M. Chitalu}

\date{}

\begin{document}
	
	\begin{abstract}
				Fixed corotated elasticity contains two fractions in its Hessian.
				One divides by the difference of two signed principal stretches, and the other divides by their sum.
				Both denominators can vanish, yet only one marks a singularity of the energy.
				Two polynomials whose coefficients come directly from the deformation gradient explain the difference.
				The characteristic cubic of the signed stretch tensor has the stretch differences as its root gaps.
				A quartic whose roots are the traces of the four signed stretch tensors has the stretch sums as the gaps between its selected root and the other three.
				Interpolation cancels the difference denominators once the stress is assembled, and the quartic derivative at the selected root collects the stretch sums that control the rotation derivative.
				On the branch that the energy selects, a zero sum occurs exactly where the nearest rotation becomes non-unique, and a short max-function argument proves that the energy has a kink there.
				Each quartic root is the stationary value of a rotation, and averaging the selected rotation with a companion gives each stretch gradient as a singular dyad.
				This yields a signed singular value decomposition whose Lagrange and Newton forms are the matrix versions of the same interpolation.
			\end{abstract}
	
	
	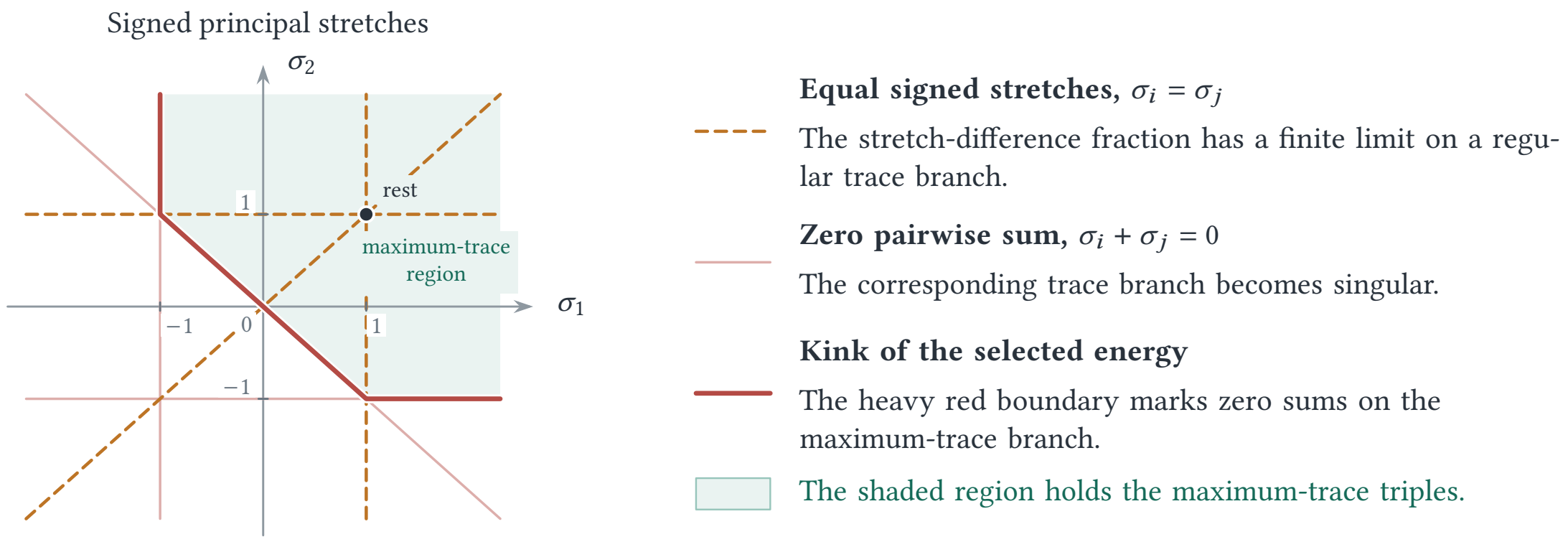
\begin{teaserfigure}
		\centering
		\begingroup
		\definecolor{stretchBlue}{HTML}{3569A8}
		\definecolor{stretchGreen}{HTML}{177D68}
		\definecolor{stretchOrange}{HTML}{BD7425}
		\definecolor{stretchRed}{HTML}{B44B45}
		\definecolor{stretchInk}{HTML}{28313B}
		\definecolor{stretchMuted}{HTML}{697580}
		\begin{tikzpicture}[
				x=0.95cm,
				y=0.85cm,
				font=\small,
				text=stretchInk,
			line cap=round,
			line join=round,
			axis/.style={
				draw=stretchMuted!75,
				line width=0.55pt,
				-{Stealth[length=1.8mm,width=1.2mm]}
			},
			coincidence/.style={
				draw=stretchOrange,
				line width=0.85pt,
				dash pattern=on 3pt off 2pt
			},
			other branch/.style={
				draw=stretchRed!45,
				line width=0.65pt
			},
			selected boundary/.style={
				draw=stretchRed,
				line width=1.25pt,
				preaction={draw=white,line width=2.3pt}
			},
			tick label/.style={
				font=\scriptsize,
				text=stretchMuted,
				fill=white,
				inner sep=1pt
			},
			legend text/.style={
				anchor=west,
				align=left,
				text width=7.0cm,
				inner sep=0pt
			}
			]
			\begin{scope}
				\node[
				font=\small,
				align=center
				] at (0.05,3.05)
				{Signed principal stretches};
				\fill[stretchGreen!9]
				(-1,1)--
				(-1,2.3)--
				(2.3,2.3)--
				(2.3,-1)--
				(1,-1)--cycle;
				\draw[axis]
				(-2.48,0)--(2.62,0)
				node[right=3pt,text=stretchInk] {$\sigma_1$};
				\draw[axis]
				(0,-2.48)--(0,2.62)
					node[right=3pt,text=stretchInk] {$\sigma_2$};
				\draw[other branch] (-2.3,2.3)--(2.3,-2.3);
				\draw[other branch] (-1,-2.3)--(-1,2.3);
				\draw[other branch] (-2.3,-1)--(2.3,-1);
				\draw[coincidence] (-2.3,-2.3)--(2.3,2.3);
				\draw[coincidence] (1,-2.3)--(1,2.3);
				\draw[coincidence] (-2.3,1)--(2.3,1);
				\draw[selected boundary]
				(-1,2.3)--
				(-1,1)--
				(1,-1)--
				(2.3,-1);
				\foreach \tick in {-1,1}{%
					\draw[
					draw=stretchMuted,
					line width=0.55pt
					] (\tick,-0.045)--(\tick,0.045);
					\node[
					tick label,
					below right
					] at (\tick,-0.07) {$\tick$};
					\draw[
					draw=stretchMuted,
					line width=0.55pt
					] (-0.045,\tick)--(0.045,\tick);	
					\node[
					tick label,
					above left
					] at (-0.07,\tick) {$\tick$};
				}
				\node[
				tick label,
				below left
				] at (-0.06,-0.06) {$0$};
				\filldraw[
				fill=stretchInk,
				draw=white,
				line width=0.7pt
				] (1,1) circle[radius=2pt];
				\node[
				anchor=south west,
				font=\scriptsize,
				text=stretchInk,
				fill=stretchGreen!9,
				inner sep=2pt
				] at (1.08,1.10) {rest};
				\node[
				text=stretchGreen!85!black,
				font=\scriptsize,
				align=center,
				fill=stretchGreen!9,
				inner sep=2pt
				] at (1.68,0.48)
				{maximum-trace\\region};
			\end{scope}
			\begin{scope}[shift={(4.20,0)}]
				\draw[coincidence]
				(0,1.90)--(0.72,1.90);
				\node[legend text] at (1.00,1.90)
				{\textbf{Equal signed stretches,} $\sigma_i=\sigma_j$\\[3pt]
						The stretch-difference fraction has a finite limit
						on a regular trace branch.};
				\draw[other branch]
				(0,0.48)--(0.72,0.48);
				\node[legend text] at (1.00,0.48)
				{\textbf{Zero pairwise sum,} $\sigma_i+\sigma_j=0$\\[3pt]
						The corresponding trace branch becomes singular.};
				\draw[selected boundary]
				(0,-0.94)--(0.72,-0.94);
				\node[legend text] at (1.00,-0.94)
				{\textbf{Kink of the selected energy}\\[3pt]
						The heavy red boundary marks zero sums
						on the maximum-trace branch.};
				\filldraw[
				fill=stretchGreen!9,
				draw=stretchGreen!45,
				line width=0.45pt
				] (0,-2.20) rectangle (0.72,-1.86);
				\node[
				legend text,
				text=stretchGreen!85!black
				] at (1.00,-2.03)
				{The shaded region holds the maximum-trace triples.};
			\end{scope}
		\end{tikzpicture}
		\endgroup
		\caption{%
				Two vanishing denominators with different meanings,
				shown in the slice $\sigma_3=1$ of signed-stretch space.
				Dashed orange lines mark equal signed stretches, where the difference fraction has a finite limit.
				Red lines mark zero pairwise sums.
				The shaded region holds the triples chosen by the maximum-trace convention (\secref{selection}),
				and its heavy red boundary marks the zero sums of the selected branch, where the energy has a kink.
				The thin pale red lines belong to other trace branches and leave the energy smooth.}
			\Description{%
			A signed-stretch plane with the third stretch fixed at one.
			Three dashed orange lines mark equal signed stretches.
			Three red lines mark zero pairwise sums.
			A pale green region identifies the maximum-trace branch.
			Its singular boundary is emphasized in heavy red:
			a vertical ray above minus one comma one,
			a diagonal segment to one comma minus one,
			and a horizontal ray extending right.
			Other portions of the zero-sum lines appear in pale red.
			The rest state is marked at one comma one.}
		\label{fig:teaser}
	\end{teaserfigure}
	
	\maketitle
	
	\section{Introduction}\label{sec:introduction}
	
	\begin{figure*}[t]
		\centering
		\begin{tikzpicture}[
			flip/.style={flipc, densely dashed, line width=1pt},
			twist/.style={twistc, line width=1pt},
			otherbranch/.style={twistc!55, dashed, line width=0.85pt},
			guide/.style={gray!55, thin},
			note/.style={font=\scriptsize, inner sep=2pt},
			]
			\begin{groupplot}[
				group style={group size=3 by 1, horizontal sep=0.055\textwidth},
				scale only axis,
				width=0.255\textwidth, height=0.17\textwidth,
				axis x line*=bottom, axis y line*=left,
				axis line style={gray!65},
				tick align=outside, tick style={gray!65},
				tick label style={font=\scriptsize},
				label style={font=\footnotesize},
				title style={font=\footnotesize\bfseries},
				every axis plot/.append style={line cap=round},
				]
				\nextgroupplot[
				title={(a) Stretch energy},
				xlabel={$s$}, ylabel={$\phi(s)$},
				xmin=-0.15, xmax=2.25, ymin=-0.18, ymax=1.68,
				xtick={0.5,1,1.5}, xticklabels={$\sigma_j$,$1$,$\sigma_i$},
				ytick={0,1},
				]
				\addplot[guide] coordinates {(-0.15,0) (2.25,0)};
				\addplot[energyc, line width=1pt, domain=-0.15:2.25, samples=81] {(x-1)^2};
				\addplot[gray!80, densely dashed, domain=0.17:0.82] {0.75-x};   %
				\addplot[gray!80, densely dashed, domain=1.18:1.83] {x-1.25};   %
				\addplot[black, only marks, mark=*, mark size=1.45pt]
				coordinates {(0.5,0.25) (1.5,0.25)};
				\node[note, anchor=south west] at (axis cs:0.28,1.26) {$\phi(s)=(s-1)^2$};
				\node[note, anchor=north east] at (axis cs:0.47,0.23) {$-1$};
				\node[note, anchor=north west] at (axis cs:1.53,0.23) {$+1$};
				\nextgroupplot[
				title={(b) Chord slopes on $\phi'$},
				xlabel={$s$}, ylabel={$y$},
				xmin=-1.65, xmax=2.45, ymin=-3.25, ymax=3.25,
				xtick={-0.5,0.5,1.5}, xticklabels={$-\sigma_j$,$\sigma_j$,$\sigma_i$},
				ytick={-2,0,2},
				]
				\addplot[guide] coordinates {(-1.65,0) (2.45,0)};
				\addplot[guide] coordinates {(0,-3.25) (0,3.25)};
				\addplot[gray!75, domain=-0.625:2.45] {2*x-2};                 %
				\addplot[gray!65, densely dashed, domain=-1.65:0.625] {2*x+2}; %
				\addplot[gray!65, densely dotted] coordinates {(0.5,-1) (-0.5,1)};
				\addplot[flip]  coordinates {(0.5,-1) (1.5,1)};
				\addplot[twist] coordinates {(-0.5,1) (1.5,1)};
				\addplot[black, only marks, mark=*, mark size=1.45pt]
				coordinates {(0.5,-1) (1.5,1) (-0.5,1)};
				\node[note, anchor=east]       at (axis cs:-0.56,1)    {$-A_j$};
				\node[note, anchor=north east] at (axis cs:0.47,-1)    {$A_j$};
				\node[note, anchor=west]       at (axis cs:1.55,1)     {$A_i$};
				\node[note, anchor=south west] at (axis cs:-1.62,1.45) {$-\phi'(-s)$};
				\node[note, anchor=south east] at (axis cs:2.22,2.45)  {$\phi'(s)$};
				\node[note, twistc, anchor=south] at (axis cs:0.48,1.10)  {$t_{ij}=0$};
				\node[note, flipc,  anchor=west]  at (axis cs:1.20,-0.93) {$l_{ij}=2$};
				\nextgroupplot[
				title={(c) Hessian quotients},
				xlabel={$\sigma_i$\quad($\sigma_j=\tfrac12$)}, ylabel={$l_{ij}$, $t_{ij}$},
				xmin=-3.2, xmax=4.5, ymin=-6, ymax=8,
				xtick={-0.5,0.5,1.5}, xticklabels={$-\tfrac12$,$\tfrac12$,$\tfrac32$},
				ytick={-4,0,2,6},
				]
				\fill[twistc!8] (axis cs:-0.5,-6) rectangle (axis cs:1.5,0);   %
				\addplot[guide] coordinates {(-3.2,0) (4.5,0)};
				\addplot[gray!75, densely dotted] coordinates {(-0.5,-6) (-0.5,8)};  %
				\addplot[otherbranch, domain=-3.2:-7/6, samples=100] {2-4/(x+0.5)};
				\addplot[twist,       domain=0:4.5,     samples=140] {2-4/(x+0.5)};
				\addplot[flip] coordinates {(-3.2,2) (4.5,2)};
				\addplot[only marks, mark=*, mark size=2.2pt,
				mark options={draw=flipc, fill=white, line width=0.8pt}]
				coordinates {(0.5,2)};
				\addplot[twistc, only marks, mark=*, mark size=1.6pt] coordinates {(1.5,0)};
				\node[note, anchor=east]                at (axis cs:-0.60,-3.05) {pole};
				\node[note, flipc, anchor=south]        at (axis cs:0.5,2.23)    {removable};
				\node[note, flipc, anchor=south east]   at (axis cs:4.45,2.18)   {$l_{ij}=2$};
				\node[note, twistc, anchor=west]        at (axis cs:1.65,-2.75)  {$t_{ij}=2-4/\varsigma$};
				\draw[->, gray!85, thin, shorten <=1pt, shorten >=1pt]
				(axis cs:3.1,2) -- (axis cs:3.1,0.888889);                    %
				\node[note, anchor=east] at (axis cs:3.07,1.46) {$-4/\varsigma$};
			\end{groupplot}
		\end{tikzpicture}
		\caption{The two Hessian fractions of \eqn{block} for $\phi(s)=(s-1)^2$.
				(a)~The tangent slopes are $-1$ at $\sigma_j=\tfrac12$ and $+1$ at $\sigma_i=\tfrac32$.
				(b)~With $A_k=(\sigma_k,\phi'(\sigma_k))$, the difference fraction $l_{ij}$ is the slope of the dashed orange chord from $A_j$ to $A_i$, here $2$. The sum fraction $t_{ij}$ is the slope of the red chord from the reflected point $-A_j$ to $A_i$, here zero.
				(c)~With $\sigma_j=\tfrac12$ fixed, the open circle marks equal stretches, where $l_{ij}$ has the limit $\phi''(\tfrac12)=2$. The sum fraction $t_{ij}=2-4/\varsigma$, with $\varsigma=\sigma_i+\sigma_j$, has a pole at opposite stretches and is negative on the shaded interval $0<\varsigma<2$. The pale dashed curve continues to $\varsigma<0$, outside the maximum-trace region.}
			\Description{Three panels show a quadratic stretch energy, its derivative and reflected derivative, and the resulting Hessian eigenvalues. The difference quotient has a removable hole at equal stretches. The sum quotient has a pole at opposite stretches.}
		\label{fig:two-quotients}
	\end{figure*}
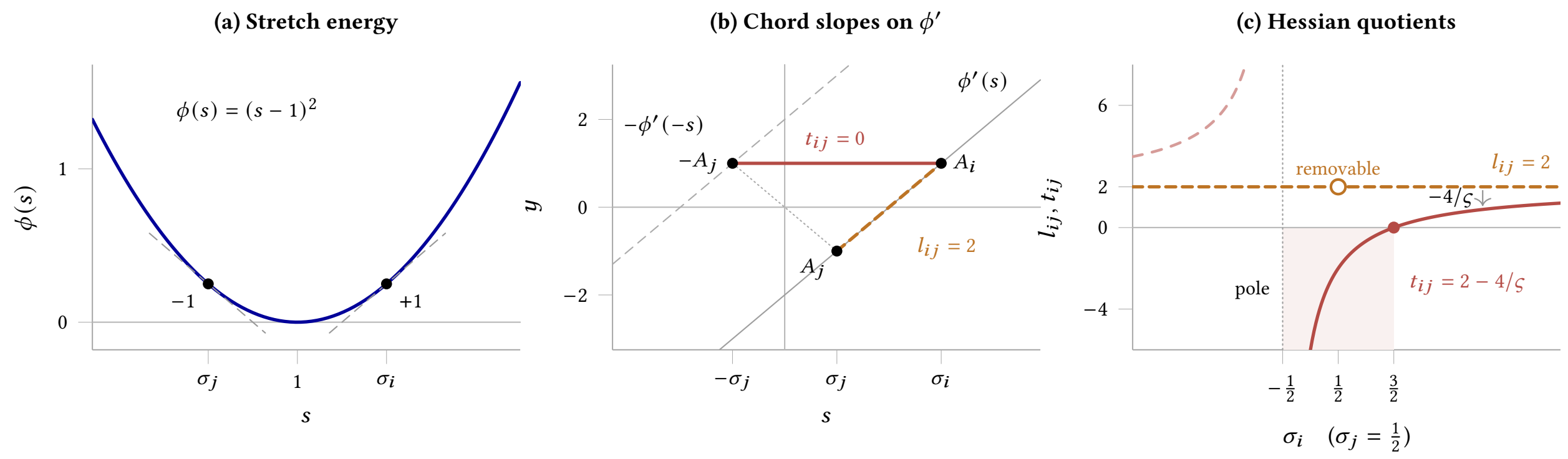
	
	A division by zero in an elasticity formula can mean two different things. The terms may cancel, so the formula has a finite limit. The energy may instead have a kink, so no limit exists. A removable division needs only a limit formula. A kink is a property of the energy itself, and how a solver treats it is a separate modelling choice. Fixed corotated elasticity \cite{Stomakhin2012}, a widely used model in graphics, contains both cases, as Figure~\ref{fig:teaser} shows.

Its Hessian in principal-stretch coordinates contains two fractions \cite{Teran2005,Stomakhin2012Supplement}, drawn in Figure~\ref{fig:two-quotients}. The first divides a difference of stress values by the difference of two signed principal stretches, and it tends to a derivative when the stretches agree. The second divides a sum of stress values by the sum of the stretches, and it grows without bound when they cancel. Established implementations already treat the two cases differently \cite{Stomakhin2012Supplement,10.1145/3532720.3535643}. This paper explains why through two polynomials whose coefficients come directly from the deformation gradient. The root gaps of a cubic are the stretch differences, and the gaps between the selected root of a trace quartic and its three companions are the stretch sums. Each denominator therefore belongs to its own polynomial.

\paragraph{Contributions.}
After the model (\secref{model}) and the two polynomials (\secref{polynomials}), the paper makes four connections explicit.
(i)~The stress coefficients are second divided differences of the stretch penalty, so coincident signed stretches need no perturbation (\secref{interpolation}).
(ii)~The selected trace root is singular exactly where the nearest proper rotation is non-unique, and a two-line argument about maxima shows that the energy has a kink there (Sections~\ref{sec:singularities} and~\ref{sec:selection}).
(iii)~The gradient of every quartic root is a stationary rotation, and the average of the selected rotation and one companion is a stretch gradient (\secref{svd-quartic}).
(iv)~The resulting signed SVD has Lagrange and Newton forms that are the matrix versions of the interpolation in~(i) (\secref{svd}), and the cubic sits inside the quartic through a single identity (Appendix~\ref{app:quartic-cubic}).
\secref{svd-evaluation} measures cost and accuracy.

\paragraph{Relation to prior work.}
The principal-stretch Hessian comes from \citet{Teran2005} and \citet{Stomakhin2012}, and \citet{Smith2019} give its analytic eigensystem. In mechanics, elastic moduli in principal axes go back to \citet{Chadwick1971}, and their evaluation at equal stretches is treated by \citet{Miehe1993,Miehe1994}, \citet{JeremicCheng2005} and \citet{Connolly2019}. The stretch tensor can be computed directly from invariants \cite{HogerCarlson1984,Franca1989,SZABO2026113971}, and the smoothness of symmetric stretch functions goes back to \citet{Ball1984}.

The trace quartic appears in several fields. In attitude determination its roots are the eigenvalues of the Davenport matrix used in QUEST \cite{shuster1981three}. \citet{Markley1993FOAM} writes the same quartic in matrix coefficients and gives an explicit rotation formula whose denominator is the product of signed stretch sums. He flags the zeros of this product as indeterminate attitudes. \citet{Horn1987} obtains the optimal rotation as the eigenvector of the largest eigenvalue of the same kind of $4\times4$ matrix. \citet{HighamNoferini2016} compute the polar decomposition from this matrix and note that its eigenvectors are the stationary points of the trace written in quaternions. They show that the four eigenvalues are signed stretch sums, a result they credit to \citet{mackey1995hamilton}, and that the derivative of the quartic at its dominant root has a particular magnitude expression in the singular values (Equation \ref{eq:alpha}).  

In graphics, \citet{Lin2022,10.1145/3666086} introduced the trace quartic to co-rotated isotropic and anisotropic elasticity, respectively. They obtained the rotation as the trace gradient $\mbR=\partial f/\partial\mbF$, the singular values from the invariants, and a closed-form Hessian through invariant derivatives. The present paper builds on that work, takes its Hessian as given, and adds the four connections above.

\citet{MyronenkoSong2009} characterize when the nearest rotation is unique, and \citet{Chao2010} and \citet{Stomakhin2012} discuss the consequences for elasticity under inversion. Rotation extraction is central to deformation methods in graphics \cite{10.5555/1281991.1282006,10.1145/2010324.1964932,10.1145/2994258.2994269,10.1145/3596711.3596731}, and the derivative and conditioning of the polar factor are well studied \cite{doi:10.1137/16M108971X,Higham2008}. Closed-form eigendecompositions of symmetric $3\times3$ matrices \cite{10.1145/355578.366316,Kopp2008Hermitian3x3} and the SVD itself \cite{Stewart1992} are classical, as are polynomial interpolation and Sylvester's formula for eigenprojectors \cite{Neumaier_2001,Higham2008}.

\section{The Model and Its Two Fractions}\label{sec:model}
	
	\begin{figure}[t]
		\centering
		\begingroup
		\definecolor{tetReference}{HTML}{3569A8}
		\definecolor{tetDeformed}{HTML}{177D68}
		\definecolor{tetInk}{HTML}{28313B}
		
		\begin{tikzpicture}[
				x=0.85cm,
				y=0.85cm,
				font=\scriptsize,
				line cap=round,
				line join=round,
				vertex label/.style={
					text=tetInk,
				inner sep=1.5pt
			},
			boundary/.style={
				line width=0.55pt
			},
			hidden edge/.style={
				line width=0.45pt,
				dash pattern=on 2pt off 1.7pt
			},
			edge vector/.style={
				line width=0.85pt,
				-{Stealth[length=1.7mm,width=1.2mm]},
				shorten <=2pt,
				shorten >=2.5pt
			},
			mapping/.style={
				draw=tetInk!75,
				line width=0.75pt,
				-{Stealth[length=2mm,width=1.4mm]}
			}
			]
			
			\begin{scope}
				\coordinate (A) at (0.35,0.00);
				\coordinate (B) at (2.05,0.12);
				\coordinate (C) at (0.90,1.65);
				\coordinate (D) at (-0.12,0.75);
				
				\fill[tetReference!8]  (A)--(B)--(C)--cycle;
				\fill[tetReference!19] (A)--(C)--(D)--cycle;
				
				\draw[hidden edge,tetReference!55]
				(B)--(D);
				\draw[boundary,tetReference!70!black]
				(A)--(B)--(C)--(D)--cycle;
				
				\draw[edge vector,tetReference]
				(A)--(B);
				\draw[edge vector,tetReference,
				preaction={draw=white,line width=2.2pt,-}]
				(A)--(C);
				\draw[edge vector,tetReference]
				(A)--(D);
				
				\foreach \P in {A,B,C,D}
				\fill[tetInk] (\P) circle[radius=1.2pt];
				
				\node[vertex label,below left]  at (A) {$\mathbf X_0$};
				\node[vertex label,below right] at (B) {$\mathbf X_1$};
				\node[vertex label,above]       at (C) {$\mathbf X_2$};
				\node[vertex label,left]        at (D) {$\mathbf X_3$};
				
				\node[text=tetReference!80!black] at (0.95,-0.48)
				{Reference edges $\mathbf D_m$};
			\end{scope}
			
			\draw[mapping]
			(2.55,0.85)--(3.65,0.85)
			node[midway,above=4pt,text=tetInk] {$\mbF$};
			
			\begin{scope}[shift={(4.10,0)}]
				\coordinate (A) at (0.35,0.00);
				\coordinate (B) at (2.25,0.10);
				\coordinate (C) at (1.55,1.65);
				\coordinate (D) at (0.28,0.85);
				
				\fill[tetDeformed!8]  (A)--(B)--(C)--cycle;
				\fill[tetDeformed!19] (A)--(C)--(D)--cycle;
				
				\draw[hidden edge,tetDeformed!55]
				(B)--(D);
				\draw[boundary,tetDeformed!70!black]
				(A)--(B)--(C)--(D)--cycle;
				
				\draw[edge vector,tetDeformed]
				(A)--(B);
				\draw[edge vector,tetDeformed,
				preaction={draw=white,line width=2.2pt,-}]
				(A)--(C);
				\draw[edge vector,tetDeformed]
				(A)--(D);
				
				\foreach \P in {A,B,C,D}
				\fill[tetInk] (\P) circle[radius=1.2pt];
				
				\node[vertex label,below left]  at (A) {$\mathbf x_0$};
				\node[vertex label,below right] at (B) {$\mathbf x_1$};
				\node[vertex label,above]       at (C) {$\mathbf x_2$};
				\node[vertex label,left]        at (D) {$\mathbf x_3$};
				
				\node[text=tetDeformed!80!black] at (1.20,-0.48)
				{Deformed edges $\mathbf D_s$};
			\end{scope}
			
		\end{tikzpicture}
		\endgroup
		\caption{%
				The deformation gradient maps the three reference edges
				to their deformed counterparts, $\mathbf D_s=\mbF\mathbf D_m$.
				Colored arrows start at vertex zero and form the columns
				of the two edge matrices.}
			\Description{%
			A blue reference tetrahedron and a green deformed tetrahedron.
			Three arrows from vertex zero identify the columns of each
			edge matrix. A central arrow labeled F indicates the mapping
			from reference edges to deformed edges. Each tetrahedron has
			two shaded visible faces and one dashed hidden edge.}
		\label{fig:tetrahedron}
	\end{figure}
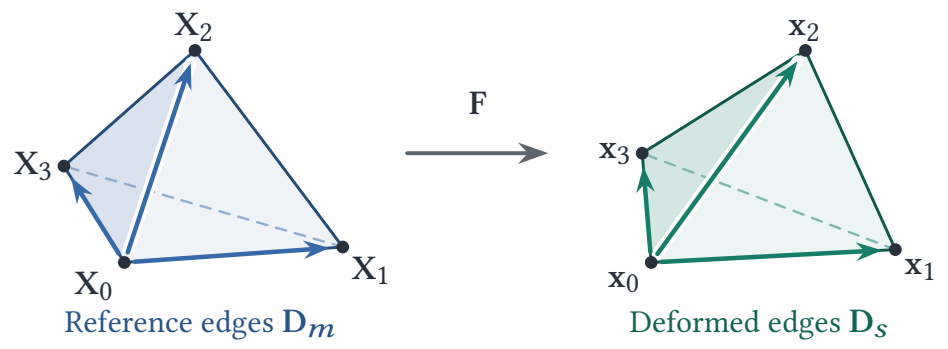
	
	\subsection{Signed stretches}
For a linear tetrahedron, the deformation gradient $\mbF=\mathbf D_s\mathbf D_m^{-1}$ maps the three reference edges to the corresponding deformed edges \cite{846276.846285,10.1145/1028523.1028541}, as Figure~\ref{fig:tetrahedron} shows. The analysis below depends only on $\mbF\in\mathbb R^{3\times3}$.

Every deformation gradient has a signed SVD and a polar decomposition,
\begin{equation}\label{eq:factors}
	\mbF=\mbU\boldsymbol\Sigma\mbV^\top=\mbR\mbS,
	\quad \mbR=\mbU\mbV^\top,
	\quad \mbS=\mbV\boldsymbol\Sigma\mbV^\top.
\end{equation}
Both $\mbU$ and $\mbV$ are proper rotations, so $\det\mbU=\det\mbV=+1$. The diagonal entries $\sigma_k$ of $\boldsymbol\Sigma$ are the signed stretches, and their magnitudes are the ordinary singular values. The rotation $\mbR$ carries the orientation of the element, and the symmetric tensor $\mbS$ carries its stretch. Negating two stretches and the matching two columns of $\mbU$ gives another valid factorization, so a sign convention is needed. The maximum-trace convention makes all stretches nonnegative when $J=\det\mbF\geq0$. For $J<0$ it negates a smallest singular value. \secref{selection} derives this convention from the energy.

We study the distortion term of fixed corotated elasticity. It measures the distance from $\mbF$ to the nearest proper rotation,
\begin{equation}\label{eq:energy}
	\psi(\mbF)=\min_{\mbR\in\SO}\|\mbF-\mbR\|_F^2
	=\sum_{k=1}^3(\sigma_k-1)^2 =\sum_k\phi(\sigma_k).
\end{equation}
The scalar penalty $\phi(s)=(s-1)^2$ acts on each stretch. The interpolation argument of \secref{interpolation} holds for any smooth $\phi$, and the rest of the paper uses this quadratic penalty.

We write $\mbP=\partial\psi/\partial\mbF$ for the first Piola--Kirchhoff stress. Its Hessian is the $9\times9$ matrix
\begin{equation}\label{eq:hessian-definition}
	\mbH=\frac{\partial^2\psi}{\partial\mbF^2}
	:=\frac{\partial\vct\mbP}{\partial(\vct\mbF)^\top}.
\end{equation}
The operator $\vct$ stacks matrix columns~\cite{Smith2018}. The gradient $\partial h/\partial\mbF$ of any scalar $h$ is a $3\times3$ matrix, and its Hessian $\mbH_h$ uses the same flattening. A prime denotes the derivative of a function of one variable, such as $\phi'$. All other derivatives use $\partial$.

\subsection{The established block structure}\label{sec:blocks}
In the principal frame, the Hessian splits into one $3\times3$ block and three $2\times2$ blocks \cite{Stomakhin2012Supplementary}. The $3\times3$ block holds second derivatives with respect to the stretches, and it is smooth wherever $\phi$ is. Each pair of indices $i\ne j$ contributes the $2\times2$ block
\begin{equation}\label{eq:block}
	\begin{aligned}
		\mathbf B_{ij}
		&=\frac{l_{ij}}2\begin{pmatrix}1&1\\1&1\end{pmatrix}
		+\frac{t_{ij}}2\begin{pmatrix}1&-1\\-1&1\end{pmatrix},\\
		l_{ij}&=\frac{\phi'(\sigma_i)-\phi'(\sigma_j)}{\sigma_i-\sigma_j},
		\qquad
		t_{ij}=\frac{\phi'(\sigma_i)+\phi'(\sigma_j)}{\sigma_i+\sigma_j}.
	\end{aligned}
\end{equation}
The block acts on the entries $(i,j)$ and $(j,i)$ of a change of $\mbF$ written in the principal frame. Equal changes in the two entries form the symmetric pair mode, with eigenvalue $l_{ij}$. Opposite changes form the antisymmetric pair mode, with eigenvalue $t_{ij}$. \citet{Smith2019} call these the flip and twist modes. Figure~\ref{fig:two-quotients} draws both fractions as chord slopes.

The chord slope $l_{ij}$ tends to $\phi''(\sigma_i)$ as $\sigma_j$ approaches $\sigma_i$. The sum fraction $t_{ij}$ has a pole where $\sigma_i+\sigma_j=0$. A merely negative $t_{ij}$ is negative curvature, which positive-semidefinite projection of the Hessian removes.

For \eqn{energy} the two eigenvalues reduce to
\begin{equation}\label{eq:arap-modes}
	l_{ij}=2,
	\qquad t_{ij}=2-\frac4{\sigma_i+\sigma_j}.
\end{equation}
The difference fraction has already cancelled, and the sum fraction keeps its pole.

\section{Two Polynomials for the Stretches}\label{sec:polynomials}
\begin{figure*}[t]
		\centering
		\begingroup
		
		\definecolor{branchGreen}{HTML}{177D68}
		\definecolor{branchBlue}{HTML}{3569A8}
		\definecolor{branchOrange}{HTML}{BD7425}
		\definecolor{branchPurple}{HTML}{98517E}
		\definecolor{figureInk}{HTML}{28313B}
		\definecolor{figureMuted}{HTML}{697580}
		\definecolor{boxEdge}{HTML}{9EAAB6}
		
		\def\DrawCubicBranch#1#2#3#4#5#6{%
			\begin{scope}[shift={(0,#1)}]
				\draw[
				-{Stealth[length=1.8mm,width=1.3mm]},
				draw=#2!85,
				line width=0.70pt
				] (8.76,0)--(9.72,0);
				
				\node[
				anchor=west,
				text=#2!85!black,
				font=\small,
				inner sep=1pt
				] at (10.00,0.34)
				{$\mathcal C(s)=#3$};
				
				\draw[
				-{Stealth[length=1.5mm,width=1.1mm]},
				draw=figureMuted!65,
				line width=0.50pt
				] (10.47,-0.10)--(17.01,-0.10);
				
				\foreach \tickNumber in {-4,-3,-2,-1,0,1,2,3,4}{%
					\draw[
					draw=boxEdge!80,
					line width=0.40pt
					]
					({13.70+0.70*\tickNumber},-0.10)
					-- ++(0,-0.055);
				}
				
				\node[
				anchor=north,
				text=figureMuted,
				font=\scriptsize,
				inner sep=1pt
				] at (13.70,-0.22) {$0$};
				
				\node[
				anchor=west,
				text=figureMuted,
				font=\scriptsize,
				inner sep=1pt
				] at (17.10,-0.10) {$s$};
				
				\foreach \stretchNumber in {#4,#5,#6}{%
					\filldraw[
					fill=#2,
					draw=white,
					line width=0.75pt
					]
					({13.70+0.70*\stretchNumber},-0.10)
					circle[radius=2.7pt];
					
					\node[
					anchor=north,
					text=#2!85!black,
					font=\scriptsize,
					inner sep=1pt
					] at ({13.70+0.70*\stretchNumber},-0.24)
					{$\stretchNumber$};
				}
			\end{scope}%
		}
		
		\resizebox{0.8\linewidth}{!}{%
				\begin{tikzpicture}[
					x=1cm,
					y=1cm,
					font=\small,
					text=figureInk,
				line cap=round,
				line join=round,
				box front/.style={
					draw=boxEdge!85,
					line width=0.50pt,
					preaction={draw=white,line width=1.40pt}
				},
				box rear/.style={
					draw=boxEdge!65,
					line width=0.45pt,
					dash pattern=on 2pt off 1.8pt
				},
				tetra front/.style={
					draw=figureMuted!90!figureInk,
					line width=0.90pt
				},
				tetra rear/.style={
					draw=figureMuted!70,
					line width=0.65pt,
					dash pattern=on 3pt off 2pt
				}
				]
				
				\node[font=\normalsize] at (8.55,6.72)
				{$\mathbf F=\operatorname{diag}(4,2,1)
					\qquad I_C=21
					\qquad II_C=273
					\qquad J=8$};
				
				\node at (8.55,6.12)
				{$\mathcal P(\zeta)
					=\zeta^4-2I_C\zeta^2-8J\zeta-I_C^2+2II_C
					=(\zeta-7)(\zeta-1)(\zeta+3)(\zeta+5)$};
				
				\node at (8.55,5.53)
				{$\displaystyle
					\mathcal P(f)=0
					\qquad
					\mathcal C(s)=s^3-fs^2+e_2s-J
					\qquad
					e_2=\frac{f^2-I_C}{2}$};
				
				\draw[boxEdge!55,line width=0.55pt]
				(-0.05,5.05)--(17.40,5.05);
				
				\node[font=\bfseries\small] at (3.70,4.58)
				{(a) Signed stretches};
				
				\node[font=\bfseries\small] at (8.60,4.58)
				{(b) Trace roots};
				
				\node[font=\bfseries\small] at (13.70,4.58)
				{(c) Stretch cubics};

				\coordinate (A) at (4.62, 3.15);  %
				\coordinate (B) at (6.94, 0.45);  %
				\coordinate (C) at (0.22,-1.35);  %
				\coordinate (D) at (3.02,-2.25);  %
				
				\coordinate (E) at (4.38, 2.25);  %
				\coordinate (F) at (7.18, 1.35);  %
				\coordinate (G) at (0.46,-0.45);  %
				\coordinate (H) at (2.78,-3.15);  %
				
				\draw[
				branchGreen!50,
				densely dotted,
				line width=0.8pt
				] (A)--(8.60,3.15);
				
				\draw[
				branchBlue!50,
				densely dotted,
				line width=0.8pt
				] (B)--(8.60,0.45);
				
				\draw[
				branchOrange!50,
				densely dotted,
				line width=0.8pt
				] (C)--(8.60,-1.35);
				
				\draw[
				branchPurple!50,
				densely dotted,
				line width=0.8pt
				] (D)--(8.60,-2.25);
				
				\fill[branchBlue!5]
				(A)--(C)--(D)--cycle;
				\fill[branchGreen!8]
				(A)--(D)--(B)--cycle;
				
				\draw[box rear] (E)--(A);
				\draw[box rear] (E)--(B);
				\draw[box rear] (E)--(C);
				
				\draw[tetra rear] (B)--(C);
				
				\draw[tetra front] (A)--(C);
				\draw[tetra front] (C)--(D);
				\draw[tetra front] (D)--(B);
				\draw[tetra front] (B)--(A);
				\draw[tetra front] (A)--(D);
				
				\draw[box front] (A)--(F);
				\draw[box front] (A)--(G);
				\draw[box front] (B)--(F);
				\draw[box front] (B)--(H);
				\draw[box front] (C)--(G);
				\draw[box front] (C)--(H);
				\draw[box front] (D)--(F);
				\draw[box front] (D)--(G);
				\draw[box front] (D)--(H);
				
				\foreach \cornerName in {E,F,G,H}{%
					\filldraw[
					fill=white,
					draw=boxEdge,
					line width=0.70pt
					] (\cornerName) circle[radius=2.0pt];
				}
				
				\foreach \cornerName/\branchColour in {A/branchGreen,B/branchBlue,C/branchOrange,D/branchPurple}{%
					\filldraw[
					fill=\branchColour,
					draw=white,
					line width=0.85pt
					] (\cornerName) circle[radius=3.3pt];
				}
				
				\node[
				anchor=south east,
				text=branchGreen,
				fill=white,
				inner sep=2pt
				] at (4.56,3.28) {$(4,2,1)$};
				
				\node[
				anchor=north east,
				text=branchBlue,
				fill=white,
				inner sep=2pt
				] at (6.84,0.30) {$(4,-2,-1)$};
				
				\node[
				anchor=south west,
				text=branchOrange,
				fill=white,
				inner sep=2pt
				] at (0.33,-1.17) {$(-4,2,-1)$};
				
				\node[
				anchor=north west,
				text=branchPurple,
				fill=white,
				inner sep=2pt
				] at (3.10,-2.43) {$(-4,-2,1)$};
				
				\node[
				text=figureMuted,
				font=\scriptsize,
				align=center
				] at (3.70,-3.83)
				{Box corners: $(\pm4,\pm2,\pm1)$\\
					Colored corners: $\sigma_1\sigma_2\sigma_3=J=8$};
				
				\draw[
				-{Stealth[length=2mm,width=1.4mm]},
				draw=figureInk!75,
				line width=0.85pt
				] (8.60,-3.45)--(8.60,3.91);
				
				\node[
				anchor=south west,
				inner sep=1pt
				] at (8.70,3.88) {$\zeta$};
				
				\draw[boxEdge,line width=0.55pt]
				(8.51,0)--(8.69,0);
				
				\node[
				anchor=east,
				text=figureMuted,
				font=\scriptsize,
				inner sep=1pt
				] at (8.39,-0.07) {$0$};
				
				\foreach \branchHeight/\traceNumber/\branchColour in {3.15/7/branchGreen,0.45/1/branchBlue,-1.35/-3/branchOrange,-2.25/-5/branchPurple}{%
					\filldraw[
					fill=\branchColour,
					draw=white,
					line width=0.8pt
					] (8.60,\branchHeight) circle[radius=3.1pt];
					
					\node[
					anchor=south east,
					text=\branchColour,
					fill=white,
					inner sep=1.5pt
					] at (8.45,{\branchHeight+0.07})
					{$f=\traceNumber$};
				}
				
				\draw[branchGreen,line width=0.7pt]
				(8.60,3.15) circle[radius=4.9pt];
				
				\node[
				anchor=south east,
				text=branchGreen,
				font=\scriptsize,
				inner sep=1pt
				] at (8.40,3.69) {maximum trace};

				\fill[
				branchGreen!5,
				rounded corners=3pt
				] (9.90,2.66) rectangle (17.35,3.86);
				
				\DrawCubicBranch{3.15}{branchGreen}
				{(s-4)(s-2)(s-1)}
				{4}{2}{1}
				
				\DrawCubicBranch{0.45}{branchBlue}
				{(s-4)(s+2)(s+1)}
				{4}{-2}{-1}
				
				\DrawCubicBranch{-1.35}{branchOrange}
				{(s+4)(s-2)(s+1)}
				{-4}{2}{-1}
				
				\DrawCubicBranch{-2.25}{branchPurple}
				{(s+4)(s+2)(s-1)}
				{-4}{-2}{1}
				
				\end{tikzpicture}%
		}
		\endgroup
		
		\caption{%
				Four signed-stretch branches of the same deformation.
				For $\mathbf F=\operatorname{diag}(4,2,1)$,
				the colored corners of the signed-stretch box have product
				$J=8$ and form a tetrahedron (a).
				Their heights are proportional to
				$\sigma_1+\sigma_2+\sigma_3$, so horizontal guides connect
				each triplet to its trace-quartic root (b).
				Each root, used as the trace $f$ of its branch, defines a cubic whose roots are that branch's stretches (c).
				Green marks the maximum-trace branch.
				Filled and hollow corners distinguish products $+8$ and $-8$,
				and dashed edges lie at the rear of the box.
				Figure~\ref{fig:stationary-rotations} shows the rotation that belongs to each root.}
			\Description{%
			Three linked panels show the signed factorizations of a
			diagonal deformation with stretches four, two, and one.
			Four colored corners of a box form a tetrahedron.
			Horizontal guides connect them to trace roots seven, one,
			minus three, and minus five.
			Each trace root leads to a factored characteristic cubic
			and a number line showing its three signed stretches.
			The maximum-trace branch is highlighted in green.}
		
		\label{fig:quartic-cubic-geometry}
	\end{figure*}
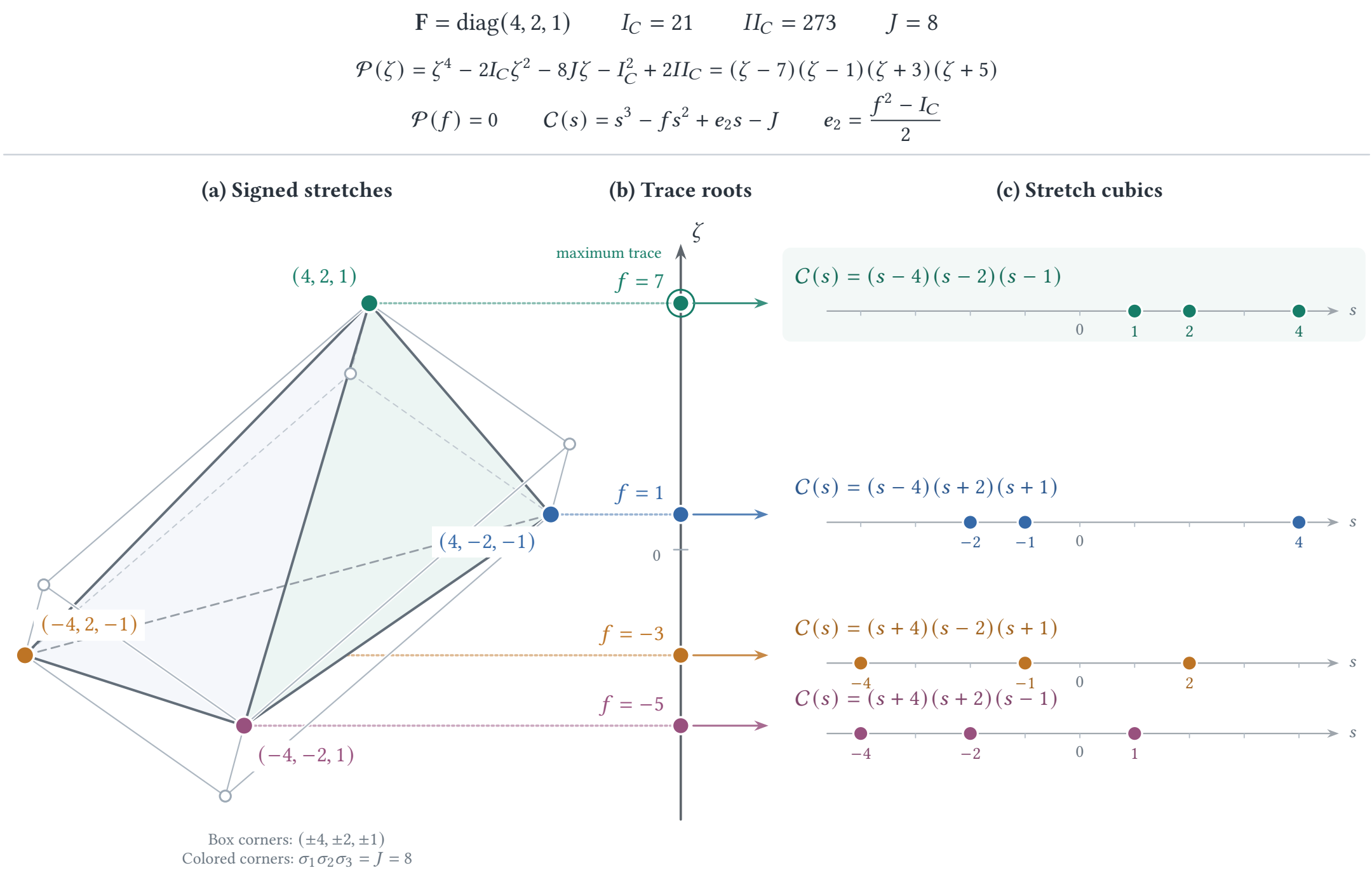

	Both fractions are written in terms of the stretches, which normally require a decomposition. Two polynomials with coefficients from $\mbF$ replace it. A quartic recovers the trace of the signed stretch tensor, and a cubic then recovers the stretches, as Figure~\ref{fig:quartic-cubic-geometry} shows for one deformation.

	\subsection{The trace quartic}
The quantities available from $\mbF$ without any decomposition are
\begin{equation}\label{eq:invariants}
	\mbC=\mbF^\top\mbF,
	\quad I_C=\tr\mbC,
	\quad II_C=\tr(\mbC^2),
	\quad J=\det\mbF.
\end{equation}
In terms of the stretches, $I_C=\sum_k\sigma_k^2$, $II_C=\sum_k\sigma_k^4$ and $J=\sigma_1\sigma_2\sigma_3$. The trace of $\mbS$ and a second symmetric function of the stretches are
\begin{equation}\label{eq:f-e2}
	f=\tr\mbS=\sum_k\sigma_k,
	\qquad e_2=\frac{f^2-I_C}{2}.
\end{equation}
The scalar $e_2$ equals $\sigma_1\sigma_2+\sigma_1\sigma_3+\sigma_2\sigma_3$. Squaring it gives $e_2^2=(I_C^2-II_C)/2+2fJ$, because the cross terms collect into $2\sigma_1\sigma_2\sigma_3(\sigma_1+\sigma_2+\sigma_3)$. Substituting $e_2=(f^2-I_C)/2$ and expanding gives the trace quartic in a placeholder variable $\zeta$,
\begin{equation}\label{eq:quartic}
	\mathcal P(\zeta)=\zeta^4-2I_C\zeta^2-8J\zeta-I_C^2+2II_C,
	\quad\mathcal P(f)=0.
\end{equation}

Negating two stretches leaves $I_C$, $II_C$ and $J$ unchanged, so each sign-flipped triple also satisfies the quartic. The four roots are therefore
\begin{equation}\label{eq:roots}
	f=\sigma_1+\sigma_2+\sigma_3,
	\qquad \zeta_k=2\sigma_k-f=\sigma_k-\sigma_i-\sigma_j,
\end{equation}
where $\{i,j,k\}=\{1,2,3\}$. The root $\zeta_k$ keeps $\sigma_k$ and negates the other two stretches. Each root is the trace of a factorization with a proper rotation, and Figure~\ref{fig:quartic-cubic-geometry}a shows the four triples as alternate corners of a box. Given $f$, the other three roots recover the stretches through $\sigma_k=(f+\zeta_k)/2$. \secref{svd} differentiates this average to obtain the singular vectors.

A monic polynomial is the product of the factors $(\zeta-\text{root})$. Its derivative at one root is therefore the product of the gaps from that root to the others. The gap from $f$ to $\zeta_k$ is $f-\zeta_k=2(\sigma_i+\sigma_j)$. Consequently
\begin{equation}\label{eq:alpha}
	\boxed{\alpha:=\mathcal P'(f)
		=8(\sigma_1+\sigma_2)(\sigma_1+\sigma_3)(\sigma_2+\sigma_3).}
\end{equation}
At a companion root the gaps are one stretch sum and two stretch differences,
\begin{equation}\label{eq:tau}
	\mathcal P'(\zeta_k)
	=-8(\sigma_i+\sigma_j)(\sigma_k-\sigma_i)(\sigma_k-\sigma_j).
\end{equation}
The selected root sees only stretch sums, so $\alpha$ vanishes exactly when a pairwise sum vanishes.

\subsection{The characteristic cubic}
The stretch tensor $\mbS$ has the signed stretches as its eigenvalues and $\det\mbS=J$. Its characteristic polynomial is
\begin{equation}\label{eq:cubic}
	\mathcal C(s)=s^3-fs^2+e_2s-J
	=\prod_{k=1}^3(s-\sigma_k).
\end{equation}
Its derivative at a distinct stretch is again a product of root gaps, now stretch differences,
\begin{equation}\label{eq:gamma}
	\gamma_k:=\mathcal C'(\sigma_k)
	=(\sigma_k-\sigma_i)(\sigma_k-\sigma_j).
\end{equation}
These are the difference factors of \eqn{tau}, and Appendix~\ref{app:quartic-cubic} places the cubic inside the quartic.

A trace branch is a root of \eqn{quartic} followed continuously as $\mbF$ changes. The branch is regular where its root is simple, which for the selected root means $\alpha\ne0$. On a regular branch the implicit function theorem makes the root a smooth function of $\mbF$. Its gradient is the corresponding rotation,
\begin{equation}\label{eq:trace-gradient}
	\frac{\partial f}{\partial\mbF}=\mbR,
	\qquad
	\frac{\partial e_2}{\partial\mbF}=f\mbR-\mbF.
\end{equation}
The second identity follows from \eqn{f-e2} and $\partial I_C/\partial\mbF=2\mbF$.

The cubic now gives the gradient of each stretch. Differentiate \eqn{cubic} with respect to $\mbF$ while holding $s$ fixed. The determinant contributes the cofactor matrix $\partial J/\partial\mbF=\cof\mbF$, which equals $J\mbF^{-\top}$ for invertible $\mbF$. Then differentiate the root equation $\mathcal C(\sigma_k)=0$, which gives $\gamma_k\,\partial\sigma_k/\partial\mbF+\partial\mathcal C/\partial\mbF=0$ at $s=\sigma_k$. The two steps give
\begin{equation}\label{eq:cubic-gradient}
	\frac{\partial\mathcal C}{\partial\mbF}
	=-s^2\mbR+s(f\mbR-\mbF)-\cof\mbF,
\end{equation}
\begin{equation}\label{eq:stretch-gradient}
	\frac{\partial\sigma_k}{\partial\mbF}
	=\frac{\sigma_k^2\mbR-\sigma_k(f\mbR-\mbF)+\cof\mbF}{\gamma_k}.
\end{equation}
Each stretch gradient divides by $\gamma_k$, a product of stretch differences, so an individual stretch can lose differentiability where two stretches meet.

\section{The Stress and its Limit}\label{sec:interpolation}
The stress is a weighted sum of stretch gradients,
\[
\mbP=\sum_{k=1}^3 \phi'(\sigma_k)\,\frac{\partial \sigma_k}{\partial \mbF}.
\]
Each term divides by stretch differences through \eqn{stretch-gradient}. This section shows that the weighted sum is a divided difference, which has a finite limit when stretches meet. The difference denominators therefore cancel before any limit is taken.

\subsection{Three shared matrix terms}
Multiply \eqn{stretch-gradient} by $\phi'(\sigma_k)$ and sum over $k$. The three matrices $\mbR$, $f\mbR-\mbF$ and $\cof\mbF$ are the same in every term. Only their scalar weights change, and these weights collect into
\begin{equation}\label{eq:Sm}
	S_m=\sum_{k=1}^3\frac{\phi'(\sigma_k)\sigma_k^m}{\gamma_k},
	\qquad m=0,1,2.
\end{equation}
Factoring out the matrices gives
\begin{equation}\label{eq:assembled}
	\boxed{\mbP=S_2\mbR-S_1(f\mbR-\mbF)+S_0\cof\mbF.}
\end{equation}
Every difference denominator now sits inside three scalar coefficients.

\subsection{The coefficients are divided differences}
Let $p_m(s)=s^m\phi'(s)$. The quadratic $\pi_m$ that passes through the three points $(\sigma_k,p_m(\sigma_k))$ has the Lagrange form
\begin{equation}\label{eq:lagrange}
	\pi_m(s)=\sum_k p_m(\sigma_k)
	\frac{(s-\sigma_i)(s-\sigma_j)}{(\sigma_k-\sigma_i)(\sigma_k-\sigma_j)}.
\end{equation}
Each numerator is a monic quadratic, so term $k$ contributes exactly $p_m(\sigma_k)/\gamma_k$ to the coefficient of $s^2$. Therefore $S_m$ is the leading coefficient of $\pi_m$.

The same quadratic has a second form built from secant slopes. For a smooth scalar function $p$, the first and second divided differences are
\begin{equation}\label{eq:divided-differences}
	\begin{aligned}
		p[a,b]&=\frac{p(b)-p(a)}{b-a},\\
		p[a,b,c]&=\frac{p[b,c]-p[a,b]}{c-a}.
	\end{aligned}
\end{equation}
The first is a slope. The second measures how that slope changes. Newton's form of the quadratic through the points at $a$, $b$ and $c$ is
\begin{equation}\label{eq:newton}
	\pi(s)=p(a)+p[a,b](s-a)+p[a,b,c](s-a)(s-b),
\end{equation}
and its leading coefficient is $p[a,b,c]$. The interpolating quadratic is unique, so the two leading coefficients agree. This gives the key identity
\begin{equation}\label{eq:Sm-dd}
	\boxed{S_m=p_m[\sigma_1,\sigma_2,\sigma_3].}
\end{equation}

For fixed corotated elasticity, $\phi'(s)=2s-2$.
The first two polynomials,
$p_0(s)=2s-2$ and $p_1(s)=2s^2-2s$, already have degree
at most two, so each equals its own interpolant.
For $p_2(s)=2s^3-2s^2$, the characteristic equation
$\mathcal C(\sigma_k)=0$ gives
$\sigma_k^3=f\sigma_k^2-e_2\sigma_k+J$ at every node.
The three interpolants are therefore
\begin{equation}\label{eq:explicit-interpolants}
	\begin{aligned}
		\pi_0(s)&=2s-2,\\
		\pi_1(s)&=2s^2-2s,\\
		\pi_2(s)&=2(f-1)s^2-2e_2s+2J.
	\end{aligned}
\end{equation}
Reading off their quadratic coefficients gives
\begin{equation}\label{eq:quadratic-S}
	S_0=0,\qquad S_1=2,\qquad S_2=2(f-1).
\end{equation}
Substitution into \eqn{assembled} recovers
$\mbP=2(\mbF-\mbR)$.

These coefficients stay finite when stretches coincide.
For a general $\phi$, repeated nodes use the
divided-difference limits
\[
p[a,a,b]=\frac{p[a,b]-p'(a)}{b-a},
\qquad
p[a,a,a]=\frac12p''(a).
\]
These limits hold when $\phi'$ is twice continuously differentiable near the nodes. On a regular trace branch the assembled stress therefore extends continuously through coincidence, even where the individual stretch gradients fail to exist.

In floating point, nearly equal nodes lose digits to cancellation, so an implementation switches to these limits below a small gap threshold, and a modest threshold suffices in simulation. The Hessian of a general $\phi$ through coincidences also needs the smoothness theory of \citet{Ball1984}. For the quadratic penalty \secref{singularities} gives it in closed form.

The only denominator left in the stress is $\alpha$, which enters through $\partial f/\partial\mbF=\mbR$.

\section{The Denominator That Remains}\label{sec:singularities}
This section links the stretch sum in $t_{ij}$ to the derivative of the rotation and then to $\alpha$.

\subsection{Rotation and invariant derivatives}
For the energy \eqn{energy}, the selected trace determines everything. Expanding the square on a regular selected branch gives
\begin{equation}\label{eq:arap-derivatives}
	\psi=I_C-2f+3,
	\quad\mbP=2(\mbF-\mbR),
	\quad\mbH=2\mbI_9-2\mbH_f.
\end{equation}
Here $\mbH_f=\partial\vct\mbR/\partial(\vct\mbF)^\top$ is the Hessian of $f$ and the derivative of the rotation. These invariant formulas are established \cite{KimEberle2022}. The stress is smooth wherever $\mbR$ is, and the Hessian is bounded wherever $\partial\mbR/\partial\mbF$ is.

Each simple root of the quartic is a smooth function of $\mbF$. For a scalar argument $z$, define
\begin{equation}\label{eq:quartic-numerator}
	\mathcal N(z)=4(z^2+I_C)\mbF-8\mbF\mbC+8z\cof\mbF.
\end{equation}
This matrix is $-\partial\mathcal P/\partial\mbF$ at fixed $z$, assembled from $\partial I_C/\partial\mbF=2\mbF$, $\partial II_C/\partial\mbF=4\mbF\mbC$ and $\partial J/\partial\mbF=\cof\mbF$. Differentiating $\mathcal P(z)=0$ while the invariants vary with $\mbF$ gives the gradient of every simple trace root,
\begin{equation}\label{eq:root-gradient}
	\frac{\partial z}{\partial\mbF}=\frac{\mathcal N(z)}{\mathcal P'(z)}.
\end{equation}
At the selected root this is the polar rotation,
\begin{equation}\label{eq:beta}
	\mbR=\frac{\mathcal N(f)}{\alpha}.
\end{equation}
Every quantity on the right comes from $\mbF$ directly, the only division is by $\alpha$, and the formula is Equation~14 of \citet{Markley1993FOAM} in elasticity notation. \secref{svd} interprets the gradients at the other roots.

\subsection{Why the stretch sums matter}\label{sec:wtssm}
The relation $\mbH=2\mbI_9-2\mbH_f$ turns each Hessian eigenvalue $h$ in \eqn{arap-modes} into the eigenvalue $(2-h)/2$ of the rotation derivative \cite{Smith2019}. The symmetric pair mode has $l_{ij}=2$ and gives zero. The antisymmetric pair mode has $t_{ij}$ and gives $2/(\sigma_i+\sigma_j)$. The three stretch modes have Hessian eigenvalue $2$ and also give zero. The largest factor by which the rotation derivative amplifies a small change of $\mbF$, measured in the Frobenius norm, is therefore
\begin{equation}\label{eq:rotation-norm}
	\left\|\frac{\partial\mbR}{\partial\mbF}\right\|_{F\to F}
	=\frac2{\min_{i<j}|\sigma_i+\sigma_j|}.
\end{equation}
This sensitivity grows without bound as a pairwise sum approaches zero.

For positive stretches this is the standard conditioning of the polar decomposition, and the same blocks give the signed case. The product $\alpha$ of \eqn{alpha} now has a precise role. It vanishes exactly at the zero pairwise sums that make the rotation derivative unbounded.

The test $\alpha=\mathcal P'(f)$ examines only the selected root. A discriminant also vanishes when two companion roots meet, which happens at equal stretches, as Appendix~\ref{app:discriminants} shows. At rest, for example,
\begin{equation}\label{eq:rest}
	\mathcal P(\zeta)=(\zeta-3)(\zeta+1)^3,
	\qquad\mathcal P'(3)=64.
\end{equation}
The three companion roots coincide, the selected root stays simple, and the rotation is smooth at rest. Only the stretch sums of the selected root affect the energy. The next section identifies that root and the configurations where its sums vanish.

\begin{figure*}[t]
			\centering
			\begingroup
\begin{tikzpicture}[
	x={(-0.437cm,-0.206cm)}, y={(0.602cm,-0.150cm)}, z={(0cm,0.700cm)},
	font=\scriptsize, line cap=round, line join=round,
	rotaxis/.style={line width=1.0pt,-{Stealth[length=1.7mm,width=1.25mm]}},
	flipaxis/.style={line width=1.0pt,dash pattern=on 1.8pt off 1.2pt,-{Stealth[length=1.7mm,width=1.25mm]}}]
\definecolor{rotInk}{HTML}{28313B}
\definecolor{rotMuted}{HTML}{697580}
\definecolor{rotTet}{HTML}{177D68}
\pgfdeclareradialshading{rotSphere}{\pgfpoint{-0.30cm}{0.32cm}}{%
	color(0cm)=(white);color(0.45cm)=(rotMuted!8);color(0.95cm)=(rotMuted!30);color(1.2cm)=(rotMuted!38)}
\newcommand{\RotPanel}[9]{%
\begin{scope}[shift={(#1cm,0cm)}]
	\fill[rotTet!9]  (0,0,0)--(4,0,0)--(0,0,1)--cycle;
	\fill[rotTet!5]  (4,0,0)--(0,2,0)--(0,0,1)--cycle;
	\draw[rotTet!50,line width=0.4pt]
		(4,0,0)--(0,2,0)--(0,0,1)--cycle
		(0,0,0)--(4,0,0) (0,0,0)--(0,2,0) (0,0,0)--(0,0,1);
	\shade[shading=rotSphere,opacity=0.80] (0,0,0) circle[radius=0.744cm];
	\draw[rotMuted!60,line width=0.45pt] (0,0,0) circle[radius=0.744cm];
	\draw[rotMuted!30,line width=0.3pt,densely dotted]
		plot[domain=114:294,samples=40,variable=\t] ({cos(\t)},0,{sin(\t)})
		plot[domain=122:302,samples=40,variable=\t] (0,{cos(\t)},{sin(\t)})
		plot[domain=126:306,samples=40,variable=\t] ({cos(\t)},{sin(\t)},0);
	\draw[rotMuted!60,line width=0.35pt]
		plot[domain=-66:114,samples=40,variable=\t] ({cos(\t)},0,{sin(\t)})
		plot[domain=-58:122,samples=40,variable=\t] (0,{cos(\t)},{sin(\t)})
		plot[domain=-54:126,samples=40,variable=\t] ({cos(\t)},{sin(\t)},0);
	\draw[rotTet!70,line width=0.4pt,dash pattern=on 1.2pt off 1pt]
		(0,0,0)--(1,0,0) (0,0,0)--(0,1,0) (0,0,0)--(0,0,1);
	\foreach \P in {(4,0,0),(0,2,0),(0,0,1)} \fill[rotTet!80!black] \P circle[radius=0.8pt];
	\ifnum#9=1 \draw[#2!60,line width=0.4pt] (1,0,0)--(2.35,0,0); \fi
	\ifnum#9=2 \draw[#2!60,line width=0.4pt] (0,1,0)--(0,1.95,0); \fi
	\ifnum#9=3 \draw[#2!60,line width=0.4pt] (0,0,1)--(0,0,1.80); \fi
	\ifnum#9=1 \draw[#2,line width=0.55pt,-{Stealth[length=1.2mm,width=1mm]}]
		plot[domain=30:320,samples=40,variable=\t] ({2.35},{0.30*cos(\t)},{0.30*sin(\t)}); \fi
	\ifnum#9=2 \draw[#2,line width=0.55pt,-{Stealth[length=1.2mm,width=1mm]}]
		plot[domain=30:320,samples=40,variable=\t] ({0.30*cos(\t)},{1.95},{0.30*sin(\t)}); \fi
	\ifnum#9=3 \draw[#2,line width=0.55pt,-{Stealth[length=1.2mm,width=1mm]}]
		plot[domain=30:320,samples=40,variable=\t] ({0.30*cos(\t)},{0.30*sin(\t)},{1.80}); \fi
	\ifnum#3>0 \draw[rotaxis,#2] (0,0,0)--(1,0,0);
	\else \draw[flipaxis,#2] (0,0,0)--(-1,0,0); \fi
	\ifnum#4>0 \draw[rotaxis,#2] (0,0,0)--(0,1,0);
	\else \draw[flipaxis,#2] (0,0,0)--(0,-1,0); \fi
	\ifnum#5>0 \draw[rotaxis,#2] (0,0,0)--(0,0,1);
	\else \draw[flipaxis,#2] (0,0,0)--(0,0,-1); \fi
	\fill[rotInk] (0,0,0) circle[radius=0.9pt];
	\node[text=#2!85!black,font=\footnotesize] at (0.25cm,1.62cm) {#6};
	\node[text=rotInk,align=center] at (0.25cm,-1.75cm)
		{$\zeta=\tr(\mathbf Q^\top\mbF)=#7$\\[1pt]$\|\mbF-\mathbf Q\|_F^2=#8$};
\end{scope}}
\RotPanel{0.0}{branchGreen}{1}{1}{1}{$\mathbf Q=\mbR=\mbI$}{4+2+1=7}{10}{0}
\node[text=rotTet!85!black,anchor=north east,inner sep=1pt] at (4,0,0) {$\mbF\mathbf e_1$};
\node[text=rotTet!85!black,anchor=west,inner sep=1.5pt] at (0,2,0) {$\mbF\mathbf e_2$};
\node[text=rotInk,anchor=north,inner sep=1.5pt,font=\tiny] at (0,0.3,-0.2) {$\mathbf x_0$};
\RotPanel{4.25}{branchBlue}{1}{-1}{-1}{$\mathbf Q_1=\mbR\mathbf H_1$}{4-2-1=1}{22}{1}
\RotPanel{8.50}{branchOrange}{-1}{1}{-1}{$\mathbf Q_2=\mbR\mathbf H_2$}{-4+2-1=-3}{30}{2}
\RotPanel{12.75}{branchPurple}{-1}{-1}{1}{$\mathbf Q_3=\mbR\mathbf H_3$}{-4-2+1=-5}{34}{3}
\end{tikzpicture}
			\endgroup
			\caption{%
				Each trace root belongs to a stationary rotation.
				For $\mbF=\operatorname{diag}(4,2,1)$, the green tetrahedron is the deformed element
				with edges $\mbF\mathbf e_j$, drawn at true scale against the unit sphere.
				Its vertex $\mathbf x_0$ sits at the sphere centre, with edges dashed inside the sphere. Dotted arcs lie behind.
				Arrows show the rotated reference axes $\mathbf Q\mathbf e_j$.
				Solid arrows point along the matching deformed edge, and dashed arrows point against it.
				The trace $\tr(\mathbf Q^\top\mbF)=\sum_j(\mathbf Q\mathbf e_j)\cdot(\mbF\mathbf e_j)$ adds each stretch with the sign of its arrow
				and gives the roots of Figure~\ref{fig:quartic-cubic-geometry}.
				The selected rotation $\mbR=\mbI$ aligns all three axes and reaches the largest root and the smallest distance $I_C+3-2\zeta$. Each $\mathbf Q_k=\mbR\mathbf H_k$ is a half turn about the $k$-th principal axis, marked by a ring.
				The average $(\mbR+\mathbf Q_k)/2$ keeps the shared arrow and cancels the two opposed arrows,
				which leaves the stretch gradient \eqn{quartic-stretch-gradient}.}
			\Description{%
				Four panels, each with a unit sphere and a green tetrahedron whose edges
				have lengths four, two and one along three axes.
				Colored arrows from the sphere center show the rotated reference axes.
				In the first panel all three arrows point along the edges and the trace is seven.
				In each other panel one arrow keeps its direction, two arrows reverse,
				a ring marks the half-turn axis, and the traces are one, minus three and minus five.}
			\label{fig:stationary-rotations}
		\end{figure*}
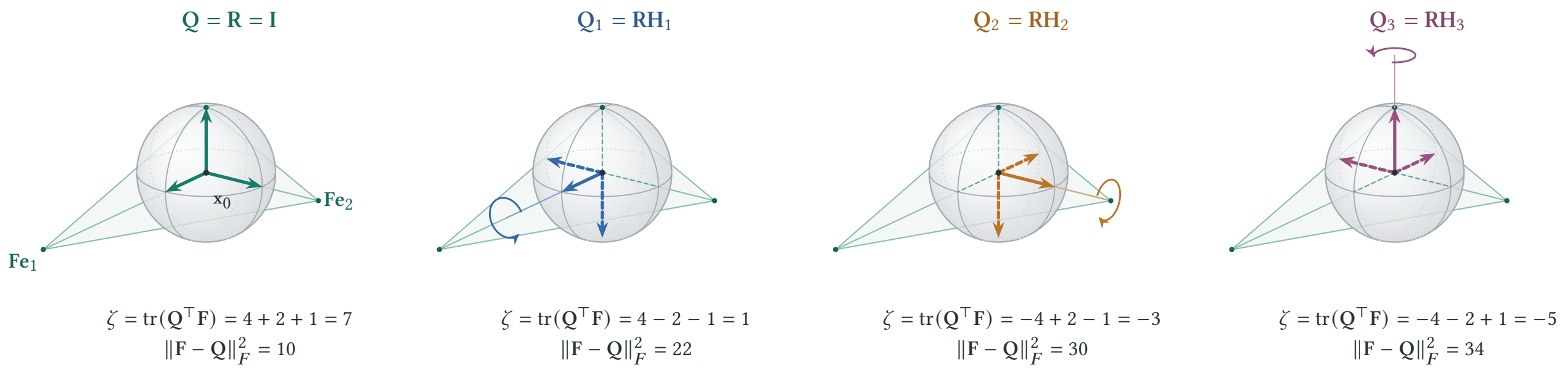
		\section{Zero Sums Are Kinks}\label{sec:selection}
Each quartic root is the trace of a signed factorization of $\mbF$. The energy uses only one of them, the one whose rotation lies nearest to $\mbF$. This section identifies that root, finds where its $\alpha$ vanishes, and shows that the energy has a kink there.

\subsection{The largest root is the model's choice}
Expanding the square gives $\|\mbF-\mbR\|_F^2=I_C-2\tr(\mbR^\top\mbF)+3$. The nearest rotation therefore maximizes a trace,
\begin{equation}\label{eq:variational}
	f=\max_{\mbR\in\SO}\tr(\mbR^\top\mbF),
	\qquad\psi=I_C+3-2f.
\end{equation}
The trace $\tr(\mbR^\top\mbF)=\sum_j(\mbR\mathbf e_j)\cdot(\mbF\mathbf e_j)$ measures how well the rotated reference axes line up with the deformed edges. Figure~\ref{fig:stationary-rotations} shows this alignment for one deformation.

At a maximizing rotation, every small turn of $\mbR$ leaves the trace unchanged to first order. Write such a turn as a path $\mbR(t)$ with $\mbR(0)=\mbR$ and $\partial\mbR/\partial t=\mbR\mathbf A$ at $t=0$. The matrix $\mathbf A$ is skew symmetric, $\mathbf A^\top=-\mathbf A$, because skew matrices are exactly the angular velocities of rotations. With $\mathbf M=\mbR^\top\mbF$, the rate of change of the trace is
\begin{equation}\label{eq:trace-rate}
	\frac{\partial}{\partial t}\tr\bigl(\mbR(t)^\top\mbF\bigr)\Big|_{t=0}
	=\tr(\mathbf A^\top\mathbf M)=-\tr(\mathbf A\mathbf M).
\end{equation}
A skew matrix times a symmetric matrix always has zero trace. Choosing $\mathbf A$ equal to the skew part of $\mathbf M$ instead gives a rate equal to the squared norm of that skew part. A maximum therefore forces the skew part to vanish, and $\mathbf M$ is symmetric.

A symmetric $\mathbf M$ with $\mbF=\mbR\mathbf M$ is a stretch tensor of the kind in \eqn{factors}. It satisfies $\mathbf M^2=\mathbf M^\top\mathbf M=\mbF^\top\mbF=\mbC$, so its eigenvalues are signed singular values. Their product is $\det\mathbf M=J$ because $\det\mbR=1$. The trace of $\mathbf M$ is therefore one of the four quartic roots. Conversely, every root has such a rotation, and Figure~\ref{fig:stationary-rotations} shows all four. The maximum in \eqn{variational} is therefore the largest root of \eqn{quartic}. With ordinary singular values $s_1\geq s_2\geq s_3\geq0$ it gives
\begin{equation}\label{eq:selected-trace}
	f=\begin{cases}s_1+s_2+s_3,&J\geq0,\\s_1+s_2-s_3,&J<0.\end{cases}
\end{equation}
At rank two the last singular value is zero, and the two expressions meet. Figure~\ref{fig:teaser} shows the resulting maximum-trace region.

		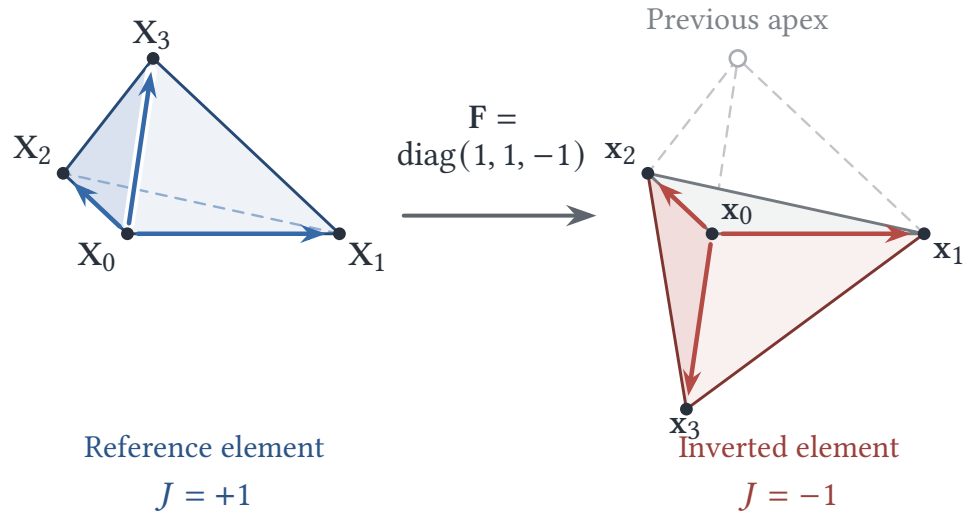
\begin{figure}[t]
		\centering
		\begingroup
		\definecolor{tetReference}{HTML}{3569A8}
		\definecolor{tetInverted}{HTML}{B44B45}
		\definecolor{tetInk}{HTML}{28313B}
		
		\begin{tikzpicture}[
				x=0.85cm,
				y=0.8cm,
				font=\scriptsize,
				line cap=round,
				line join=round,
				vertex label/.style={
				text=tetInk,
				inner sep=1.5pt
			},
			boundary/.style={
				line width=0.55pt
			},
			hidden edge/.style={
				line width=0.45pt,
				dash pattern=on 2pt off 1.7pt
			},
			edge vector/.style={
				line width=0.85pt,
				-{Stealth[length=1.7mm,width=1.2mm]},
				shorten <=2pt,
				shorten >=2.5pt
			},
			mapping/.style={
				draw=tetInk!75,
				line width=0.75pt,
				-{Stealth[length=2mm,width=1.4mm]}
			},
			ghost edge/.style={
				draw=tetInk!28,
				line width=0.45pt,
				dash pattern=on 2pt off 1.7pt
			}
			]
			
			\begin{scope}
				\coordinate (A) at (0.00,0.00);
				\coordinate (B) at (1.65,0.00);
				\coordinate (C) at (-0.50,0.50);
				\coordinate (D) at (0.20,1.45);
				
				\fill[tetReference!8]  (A)--(B)--(D)--cycle;
				\fill[tetReference!19] (A)--(D)--(C)--cycle;
				
				\draw[hidden edge,tetReference!55]
				(B)--(C);
				\draw[boundary,tetReference!70!black]
				(A)--(B)--(D)--(C)--cycle;
				
				\draw[edge vector,tetReference]
				(A)--(B);
				\draw[edge vector,tetReference]
				(A)--(C);
				\draw[edge vector,tetReference,
				preaction={draw=white,line width=2.2pt,-}]
				(A)--(D);
				
				\foreach \P in {A,B,C,D}
				\fill[tetInk] (\P) circle[radius=1.2pt];
				
				\node[vertex label,below left]  at (A) {$\mathbf X_0$};
				\node[vertex label,below right] at (B) {$\mathbf X_1$};
				\node[vertex label,above left]  at (C) {$\mathbf X_2$};
				\node[vertex label,above]       at (D) {$\mathbf X_3$};
				
				\node[
				text=tetReference!80!black,
				align=center
				] at (0.60,-1.98)
				{Reference element\\[2pt]$J=+1$};
			\end{scope}
			
			\draw[mapping]
			(2.15,0.15)--(3.65,0.15);
			
			\node[
			text=tetInk,
			align=center
			] at (2.84,0.78)
				{$\mbF=$\\$\operatorname{diag}(1,1,-1)$};
			
			\begin{scope}[shift={(4.55,0)}]
				\coordinate (A)    at (0.00,0.00);
				\coordinate (B)    at (1.65,0.00);
				\coordinate (C)    at (-0.50,0.50);
				\coordinate (D)    at (-0.20,-1.45);
				\coordinate (Dold) at (0.20,1.45);
				
				\draw[ghost edge] (A)--(Dold);
				\draw[ghost edge] (B)--(Dold)--(C);
				\draw[
				draw=tetInk!40,
				fill=white,
				line width=0.5pt
				] (Dold) circle[radius=1.5pt];
				
				\node[
				text=tetInk!60,
				above=3pt,
				inner sep=1pt
				] at (Dold) {Previous apex};
				
				\fill[tetInverted!8]  (A)--(B)--(D)--cycle;
				\fill[tetInverted!19] (A)--(D)--(C)--cycle;
				
				\fill[tetInk!6] (A)--(B)--(C)--cycle;
				
				\draw[boundary,tetInverted!70!black]
				(B)--(D)--(C);
				\draw[boundary,tetInk!65]
				(A)--(B)--(C)--cycle;
				
				\draw[edge vector,tetInverted]
				(A)--(B);
				\draw[edge vector,tetInverted]
				(A)--(C);
				\draw[edge vector,tetInverted]
				(A)--(D);
				
				\foreach \P in {A,B,C,D}
				\fill[tetInk] (\P) circle[radius=1.2pt];
				
				\node[vertex label,above right] at (A) {$\mathbf x_0$};
				\node[vertex label,below right] at (B) {$\mathbf x_1$};
				\node[vertex label,above left]  at (C) {$\mathbf x_2$};
				\node[vertex label,below]       at (D) {$\mathbf x_3$};
				
				\node[
				text=tetInverted!85!black,
				align=center
				] at (0.60,-1.98)
				{Inverted element\\[2pt]$J=-1$};
			\end{scope}
			
		\end{tikzpicture}
		\endgroup
		\caption{%
				Orientation reversal across a fixed base triangle.
				The deformation $\mbF=\operatorname{diag}(1,1,-1)$
				moves vertex~3 to the other side of the base and gives
				$J=-1$.
				The faint outline marks its previous position.
				All three singular values equal one,
				and the maximum-trace convention assigns signed stretches
				$(1,1,-1)$.
				The two smallest magnitudes are equal and $J<0$,
				so this configuration lies in the singular set of \eqn{singular-set}
				and its nearest proper rotation is non-unique.}
			\Description{%
			A blue reference tetrahedron and a red inverted tetrahedron.
			The three base vertices remain fixed, while the apex moves
			to the opposite side of their plane.
			A faint dashed outline shows the original apex position
				above the base.
			Colored arrows identify the three edges from vertex zero.
			The central mapping has diagonal entries one, one, and minus one.}
		\label{fig:inverted-tetrahedron}
	\end{figure}

		\subsection{Where the selected sums vanish}
Substituting the selected signs into \eqn{alpha} yields
\begin{equation}\label{eq:selected-alpha}
	\frac\alpha8=
	\begin{cases}
		(s_1+s_2)(s_1+s_3)(s_2+s_3),&J\geq0,\\
		(s_1+s_2)(s_1-s_3)(s_2-s_3),&J<0.
	\end{cases}
\end{equation}
For $J\geq0$ this product vanishes exactly when $s_2=s_3=0$. For $J<0$ all singular values are positive, and the product vanishes exactly when $s_2=s_3$. Figure~\ref{fig:inverted-tetrahedron} shows one such element. This proves the selected-branch characterization
\begin{equation}\label{eq:singular-set}
	\alpha=0\ \Longleftrightarrow\
	\left\{\begin{array}{l}
		\operatorname{rank}\mbF\leq1,\quad\text{or}\\
		J<0\ \text{and}\ s_2=s_3.
	\end{array}\right.
\end{equation}
An ordinary rank-two flattening stays smooth. For signed stretches $(a,b,0)$ with $a,b>0$ we have $\alpha=8ab(a+b)>0$. Equal positive stretches alone also leave $\alpha$ positive.

This set is exactly where the nearest proper rotation is non-unique. At rank one the null plane of $\mbF$ leaves a free turn, and at rank zero every rotation is a maximizer. Under inversion with $s_2=s_3$, any direction in the plane of the two smallest stretches can take the negative sign. Conversely, the maximizer is unique whenever $\operatorname{rank}\mbF\geq2$ and, for $J<0$, the smallest singular value is simple \cite{MyronenkoSong2009}.

\paragraph{A tie is a kink.}
Write $\mathcal M$ for the set of maximizing rotations in \eqn{variational}. Danskin's theorem \cite{Danskin1967} gives the directional derivative of the maximum along a direction $\boldsymbol\Delta$ as the largest rate among the maximizers:
\begin{equation}\label{eq:danskin}
	\begin{aligned}
		\lim_{t\to0^+}\frac{f(\mbF+t\boldsymbol\Delta)-f(\mbF)}{t}
		&=\lim_{t\to0^+}\frac{1}{t}\Big[
		\max_{\mathbf Q\in SO(3)}
		\tr\bigl(\mathbf Q^\top(\mbF+t\boldsymbol\Delta)\bigr)\\
		&\qquad\qquad\qquad
		-\max_{\mathbf Q\in SO(3)}
		\tr(\mathbf Q^\top\mbF)\Big]\\
		&=\max_{\mathbf Q\in\mathcal M}
		\tr(\mathbf Q^\top\boldsymbol\Delta).
	\end{aligned}
\end{equation}
If the maximizer is unique, say $\mbR$, this reduces to $\tr(\mbR^\top\boldsymbol\Delta)$, the inner product of $\mbR$ with $\boldsymbol\Delta$. Thus $f$ is differentiable, with gradient $\partial f/\partial\mbF=\mbR$.

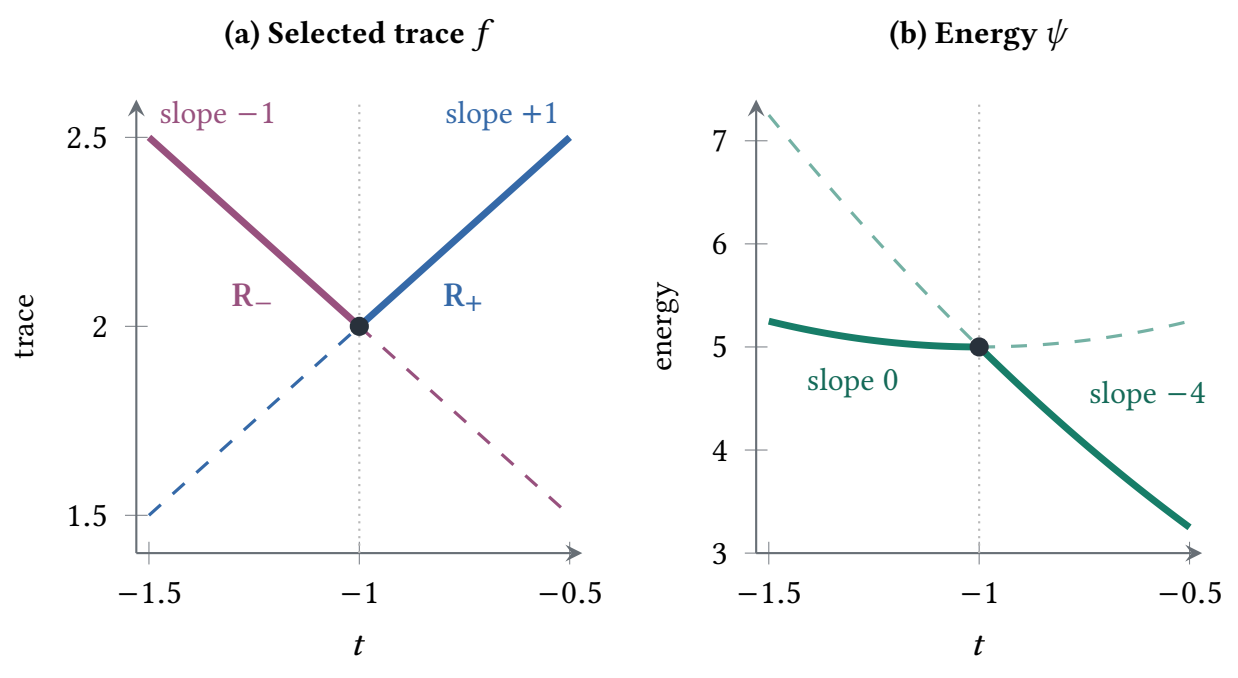
\begin{figure}[t]
	\centering
	\begingroup
	\begin{tikzpicture}
		\definecolor{kinkBlue}{HTML}{3569A8}
		\definecolor{kinkPurple}{HTML}{98517E}
		\definecolor{kinkGreen}{HTML}{177D68}
		\definecolor{kinkInk}{HTML}{28313B}
		\begin{groupplot}[
			group style={group size=2 by 1, horizontal sep=1.15cm},
			scale only axis, width=2.95cm, height=3.0cm,
			xmin=-1.53, xmax=-0.47, xtick={-1.5,-1,-0.5}, xlabel={$t$},
			/tikz/font=\scriptsize,
			axis lines=left, axis line style={draw=kinkInk!70}, tick style={draw=kinkInk!60},
			title style={font=\scriptsize\bfseries}, label style={font=\scriptsize},
			tick label style={font=\scriptsize}, clip=false, samples=61]
			\nextgroupplot[title={(a) Selected trace $f$}, ylabel={trace}, ymin=1.4, ymax=2.6, ytick={1.5,2,2.5}]
			\addplot[gray!55,densely dotted] coordinates {(-1,1.4) (-1,2.6)};
			\addplot[kinkBlue,dashed,line width=0.6pt,domain=-1.5:-1] {x+3};
			\addplot[kinkPurple,dashed,line width=0.6pt,domain=-1:-0.5] {1-x};
			\addplot[kinkPurple,line width=1.3pt,domain=-1.5:-1] {1-x};
			\addplot[kinkBlue,line width=1.3pt,domain=-1:-0.5] {x+3};
			\addplot[only marks,mark=*,mark size=1.6pt,kinkInk] coordinates {(-1,2)};
			\node[kinkPurple,anchor=south west,inner sep=1pt] at (axis cs:-1.49,2.50) {slope $-1$};
			\node[kinkBlue,anchor=south east,inner sep=1pt] at (axis cs:-0.51,2.50) {slope $+1$};
			\node[kinkPurple,anchor=north,inner sep=1pt] at (axis cs:-1.25,2.13) {$\mbR_-$};
			\node[kinkBlue,anchor=north,inner sep=1pt] at (axis cs:-0.75,2.13) {$\mbR_+$};
			\nextgroupplot[title={(b) Energy $\psi$}, ylabel={energy}, ymin=3, ymax=7.4, ytick={3,4,5,6,7}]
			\addplot[gray!55,densely dotted] coordinates {(-1,3) (-1,7.4)};
			\addplot[kinkGreen!60,dashed,line width=0.6pt,domain=-1:-0.5] {(x+1)^2+5};
			\addplot[kinkGreen!60,dashed,line width=0.6pt,domain=-1.5:-1] {(x-1)^2+1};
			\addplot[kinkGreen,line width=1.3pt,domain=-1.5:-1] {(x+1)^2+5};
			\addplot[kinkGreen,line width=1.3pt,domain=-1:-0.5] {(x-1)^2+1};
			\addplot[only marks,mark=*,mark size=1.6pt,kinkInk] coordinates {(-1,5)};
			\node[kinkGreen!85!black,anchor=north,inner sep=1.5pt] at (axis cs:-1.3,4.88) {slope $0$};
			\node[kinkGreen!85!black,anchor=south west,inner sep=1.5pt] at (axis cs:-0.76,4.3) {slope $-4$};
		\end{groupplot}
	\end{tikzpicture}
	\endgroup
	\caption{%
		A tie between two rotations creates a kink.
		Along $\mbF(t)=\operatorname{diag}(t,2,1)$ the flipped edge $t\,\mathbf e_1$ outgrows the edge $\mathbf e_3$ at $t=-1$.
		(a)~The rotations $\mbR_+=\mbI$ and $\mbR_-=\operatorname{diag}(-1,1,-1)$ give traces $t+3$ and $1-t$.
		The selected trace $f=2+|t+1|$ is the larger of the two, so the nearest rotation switches from $\mbR_-$ to $\mbR_+$ at the tie, and the slope of $f$ jumps from $-1$ to $+1$.
		(b)~The energy $\psi=I_C+3-2f=t^2+4-2|t+1|$ stays continuous, and its slope jumps from $0$ to $-4$.
		Solid segments are selected, and dashed segments are their unselected extensions.}
	\Description{%
		Two graphs along a diagonal deformation path.
		The selected trace is the upper envelope of two straight lines, which cross at t equals minus one.
		The energy is the lower envelope of two parabolas with a corner at the same point.
		Solid segments show selected branches, and dashed segments show unselected extensions.}
	\label{fig:trace-selection-kink}
\end{figure}

Now suppose there are two distinct tied maximizers, $\mathbf Q_+$ and $\mathbf Q_-$. Consider the direction connecting them, $\boldsymbol\Delta=\mathbf Q_+-\mathbf Q_-$. Along this direction, Danskin's formula gives
\[
\begin{aligned}
	f'(\mbF;\boldsymbol\Delta)
	&\geq \tr(\mathbf Q_+^\top\boldsymbol\Delta)\\
	&=3-\tr(\mathbf Q_+^\top\mathbf Q_-)>0.
\end{aligned}
\]
The inequality is strict because distinct rotations satisfy $\tr(\mathbf Q_+^\top\mathbf Q_-)<3$. Reversing the direction gives the same result, now using $\mathbf Q_-$:
\[
\begin{aligned}
	f'(\mbF;-\boldsymbol\Delta)
	&\geq \tr(\mathbf Q_-^\top(-\boldsymbol\Delta))\\
	&=3-\tr(\mathbf Q_-^\top\mathbf Q_+)>0.
\end{aligned}
\]
Thus $f$ increases in both directions along the same line (see Figure~\ref{fig:trace-selection-kink}(a), at $t=-1\pm\epsilon$). This is impossible for a differentiable function: its directional derivatives along opposite directions must be negatives of one another. Hence $f$ has no gradient at $\mbF$, and $\psi=I_C+3-2f$ has a kink.

Together with the implicit function theorem on regular branches, this proves
\begin{equation}\label{eq:kink}
	\boxed{\alpha=0\ \Longleftrightarrow\ \psi\ \text{has a kink at}\ \mbF.}
\end{equation}

In Figure~\ref{fig:trace-selection-kink}, along $\mbF(t)=\operatorname{diag}(t,2,1)$ the nearest rotation switches from $\mbI$ to a half turn at $t=-1$, where every rotation about the second axis ties and the selected trace has a corner.

The branch qualification also matters. The signed triple $(-2,1,2)$ has a zero sum, so its trace branch is singular. The maximum-trace factorization of the same deformation uses $(2,2,-1)$ and has $\alpha=32$. One deformation can therefore have a singular companion branch and a smooth selected energy. Appendix~\ref{app:discriminants} shows that a discriminant mixes the two cases.

	\section{From Trace Roots to a Signed SVD}\label{sec:svd}
Every quartic root belongs to a stationary rotation, averaging two such rotations gives a stretch gradient, and the stretch gradients deliver the singular vectors. Their polynomial forms return to the interpolation of \secref{interpolation}.

\subsection{Stretch gradients from trace roots}\label{sec:svd-quartic}
Assume first that all four trace roots are simple. Label the three companion roots by their stretches, as in \eqn{roots}, and write their gradients as
\begin{equation}\label{eq:stationary-root-gradients}
	\mbR=\frac{\partial f}{\partial\mbF},
	\qquad
	\mbR_k=\frac{\partial\zeta_k}{\partial\mbF}.
\end{equation}
\eqn{root-gradient} evaluates all four gradients from the same invariants.

\paragraph{Each root belongs to a stationary rotation.}
Let $\mbv_k$ be a unit eigenvector of $\mbS$, so $\mbS\mbv_k=\sigma_k\mbv_k$. The matrix
\begin{equation}\label{eq:relative-half-turn}
	\mathbf H_k=2\mbv_k\mbv_k^\top-\mbI
\end{equation}
keeps $\mbv_k$ and reverses every vector perpendicular to it, so it is a half turn about $\mbv_k$. It shares its eigenvectors with $\mbS$, so $\mathbf H_k\mbS$ is symmetric. The proper rotation $\mathbf Q_k=\mbR\mathbf H_k$ therefore satisfies
\begin{equation}\label{eq:stationary-root-trace}
	\mathbf Q_k^\top\mbF=\mathbf H_k\mbS,
	\qquad
	\tr(\mathbf Q_k^\top\mbF)=\sigma_k-\sigma_i-\sigma_j=\zeta_k.
\end{equation}
The matrix $\mathbf Q_k^\top\mbF$ is symmetric, so by the argument of \secref{selection} the trace is stationary at $\mathbf Q_k$. Its stationary value is the root $\zeta_k$. Figure~\ref{fig:stationary-rotations} draws $\mbR$ and the three rotations $\mathbf Q_k$ for one deformation.

\paragraph{The gradient of a root is its rotation.}
Let $\mbF(t)$ be any smooth path of deformations, and let $\mathbf Q_k(t)$ be the stationary rotation that follows it. The root along the path is $\zeta_k(t)=\tr(\mathbf Q_k(t)^\top\mbF(t))$, and the product rule gives two terms,
\begin{equation}\label{eq:stationary-root-differential}
	\frac{\partial\zeta_k}{\partial t}
	=\tr\Bigl(\mathbf Q_k^\top\frac{\partial\mbF}{\partial t}\Bigr)
	+\tr\Bigl(\frac{\partial\mathbf Q_k^\top}{\partial t}\mbF\Bigr).
\end{equation}
The second term is the change in trace caused by turning the rotation, and stationarity makes it zero. Indeed $\partial\mathbf Q_k/\partial t=\mathbf Q_k\mathbf A$ with $\mathbf A$ skew symmetric, so the second term equals $-\tr(\mathbf A\mathbf Q_k^\top\mbF)$. This pairs a skew matrix with a symmetric one and vanishes. Only the first term remains, so the gradient of the root is the rotation itself,
\begin{equation}\label{eq:root-half-turn}
	\mbR_k=\mathbf Q_k=\mbR(2\mbv_k\mbv_k^\top-\mbI).
\end{equation}

\paragraph{Averaging rotations gives a singular dyad.}
Differentiating $\sigma_k=(f+\zeta_k)/2$ now gives
\begin{equation}\label{eq:quartic-stretch-gradient}
	\boxed{\frac{\partial\sigma_k}{\partial\mbF}
		=\frac{\mbR+\mbR_k}{2}
		=\mbR\mbv_k\mbv_k^\top
		=\mathbf u_k\mbv_k^\top,}
\end{equation}
where $\mathbf u_k=\mbR\mbv_k$. This is the central identity of the section. The rotations $\mbR$ and $\mbR_k$ act identically on $\mbv_k$ and oppositely on the plane perpendicular to it. Their average keeps the direction $\mbv_k$ and cancels the other two, as the arrows of Figure~\ref{fig:stationary-rotations} show.

Multiplication by the selected rotation isolates the right singular projector,
\begin{equation}\label{eq:quartic-svd-projector}
	\mathbf L_k=\mbR^\top\frac{\partial\sigma_k}{\partial\mbF}
	=\frac{\mbI+\mbR^\top\mbR_k}{2}
	=\mbv_k\mbv_k^\top.
\end{equation}

\subsection{Root-gradient form}\label{sec:svd-root}
\paragraph{Distinct stretches.}
Assume $\alpha\ne0$ and distinct signed stretches. Compute the four roots of \eqn{quartic}, take the largest as $f$, and set $\sigma_k=(f+\zeta_k)/2$. Evaluate $\mbR=\mathcal N(f)/\alpha$ and each companion gradient $\mbR_k=\mathcal N(\zeta_k)/\mathcal P'(\zeta_k)$ from \eqn{root-gradient}. The projector \eqn{quartic-svd-projector} follows without forming $\mbS$, and its normalized largest column is $\mbv_k$. Two such directions suffice, taken from the two roots with the largest $|\mathcal P'(\zeta_k)|$. The cross product of \eqn{newton-frame-completion} completes the frame with $\det\mbV=1$, and $\mbU=\mbR\mbV$. Every division is by a quartic derivative.

\paragraph{Repeated stretches.}
\eqn{tau} shows where this route stops. On a regular selected branch every pairwise sum is positive, so $\mathcal P'(\zeta_k)$ vanishes exactly when $\sigma_k$ equals another stretch. Two equal stretches $\sigma_i=\sigma_j$ give equal companion roots $\zeta_i=\zeta_j$, and both derivatives vanish. The numerators vanish with them, so \eqn{root-gradient} takes the form $0/0$. A half turn about any axis in the repeated-stretch plane is then a stationary rotation with trace $\zeta_i$, so $\mbR_i=\partial\zeta_i/\partial\mbF$ has no single value, and the singular directions of the pair are free. The isolated root $\zeta_k$ stays simple. Its gradient gives $\mbv_k$, and any orthonormal pair in the plane perpendicular to $\mbv_k$ completes $\mbV$, as in \eqn{newton-repeated-direction}. At a triple coincidence every half turn is stationary, and $\mbV=\mbI$ is valid. 

\subsection{Lagrange polynomial form}\label{sec:svd-lagrange}
\paragraph{Distinct stretches.}
Assume $\alpha\ne0$ and distinct signed stretches. Obtain $f$, the stretches and $\mbR$ as in \secref{svd-root}, and set $\mbS=\mbR^\top\mbF$. The cubic gives a second expression for the same projector. Multiplying \eqn{stretch-gradient} by $\mbR^\top$ gives
\begin{equation}\label{eq:svd-projector}
	\begin{aligned}
		\mathbf L_k&=\frac{\mbI+\mbR^\top\mbR_k}{2}\\
		&=\frac{(\mbS-\sigma_i\mbI)(\mbS-\sigma_j\mbI)}
		{(\sigma_k-\sigma_i)(\sigma_k-\sigma_j)}.
	\end{aligned}
\end{equation}
The first line reads the singular line as the fixed axis of a relative half turn, and the second as a Lagrange polynomial that is one at $\sigma_k$ and zero at the other stretches.
To obtain the second line, use $\mbR^\top\mbF=\mbS$ and
\begin{equation}\label{eq:cofactor-identity}
	\mbR^\top\cof\mbF=\mbS^2-f\mbS+e_2\mbI.
\end{equation}
Two facts give this identity for every $\mbF$. The cofactor of a product is the product of cofactors, and $\cof\mbR=\mbR$ for a rotation, so $\mbR^\top\cof\mbF=\cof\mbS$. The Cayley--Hamilton theorem in adjugate form, $\operatorname{adj}\mbS=\mbS^2-(\tr\mbS)\mbS+e_2\mbI$, is a polynomial identity valid for every $3\times3$ matrix, and $\cof\mbS=\operatorname{adj}\mbS$ because $\mbS$ is symmetric.
The numerator then becomes
$\mbS^2-(f-\sigma_k)\mbS+(e_2-f\sigma_k+\sigma_k^2)\mbI$.
Its coefficients are the sum and the product of the other two stretches, so it factors as shown.

\begin{figure}[t]
		\centering
		\begingroup
		\definecolor{projBlue}{HTML}{3569A8}
		\definecolor{projGreen}{HTML}{177D68}
		\definecolor{projInk}{HTML}{28313B}
		
		\begin{tikzpicture}[
				x=0.9cm,
				y=0.82cm,
				font=\scriptsize,
				line cap=round,
				line join=round,
				panel title/.style={
				font=\scriptsize\bfseries,
				text=projInk,
				inner sep=1pt
			},
			annotation/.style={
				text=projInk,
				align=center,
				inner sep=1pt
			},
			vector/.style={
				line width=0.85pt,
				-{Stealth[length=1.7mm,width=1.2mm]}
			},
			alternate vector/.style={
				draw=projInk!55,
				line width=0.65pt,
				dash pattern=on 2pt off 1.5pt,
				-{Stealth[length=1.5mm,width=1.1mm]}
			},
			guide/.style={
				draw=projInk!45,
				line width=0.5pt,
				dash pattern=on 2pt off 1.5pt
			}
			]
			
			\begin{scope}
				\node[panel title] at (1.20,2.30)
				{One distinct stretch};
				
				\coordinate (O) at (0.25,0.25);
				\coordinate (E) at (0.45,1.70);
				\coordinate (P) at (1.074,0.868);
				
				\draw[
				draw=projBlue,
				line width=0.7pt
				] (0.05,0.10)--(2.13,1.66);
				
				\node[
				text=projBlue!85!black,
				above left,
				inner sep=2pt
				] at (2.13,1.66)
				{$\operatorname{span}(\mathbf v_k)$};
				
				\draw[vector,draw=projInk!60]
				(O)--(E)
				node[above left,text=projInk,inner sep=2pt]
				{$\mathbf e$};
				
				\draw[guide] (E)--(P);
				
				\draw[
				draw=projInk!55,
				line width=0.45pt
				]
				(1.170,0.940)--
				(1.098,1.036)--
				(1.002,0.964);
				
				\draw[vector,draw=projGreen]
				(O)--(P);
				
				\node[
				text=projGreen!85!black,
				below right,
				inner sep=2pt
				] at (0.74,0.62)
				{$\mathbf L_k\mathbf e$};
				
				\fill[projInk] (O) circle[radius=1.1pt];
				\fill[projGreen] (P) circle[radius=1.1pt];
				
				\node[annotation] at (1.20,-0.62)
				{Project onto the singular line.\\[2pt]
					Normalize to obtain $\mathbf v_k$.};
			\end{scope}
			
			\begin{scope}[shift={(3.75,0)}]
				\node[panel title] at (1.25,2.30)
				{Two equal stretches};
				
				\begin{scope}[
					shift={(1.25,0.85)},
					x={(0.95cm,0.15cm)},
					y={(-0.30cm,0.80cm)}
					]
					\filldraw[
					fill=projBlue!8,
					draw=projBlue!60,
					line width=0.55pt
					]
					(-1.15,-1.10)--
					( 1.15,-1.10)--
					( 1.15, 1.10)--
					(-1.15, 1.10)--cycle;
					
					\draw[alternate vector]
					(0,0)--({cos(45)},{sin(45)});
					\draw[alternate vector]
					(0,0)--({-sin(45)},{cos(45)});
					
					\draw[vector,draw=projGreen]
					(0,0)--(1,0)
					node[
					right,
					text=projGreen!85!black,
					inner sep=2pt
					] {$\mathbf v_1$};
					
					\draw[vector,draw=projGreen]
					(0,0)--(0,1)
					node[
					above,
					text=projGreen!85!black,
					inner sep=2pt
					] {$\mathbf v_2$};
					
					\fill[projInk] (0,0) circle[radius=1.1pt];
				\end{scope}
				
				\node[annotation] at (1.90,-0.62)
				{Project onto the plane.\\[2pt]
					Choose an orthonormal basis.};
			\end{scope}
			
		\end{tikzpicture}
		\endgroup
		\caption{%
			Projectors identify singular subspaces.
			A distinct stretch gives a line.
				Projecting a vector onto
				that line and normalizing recovers a singular direction.
			Two equal stretches give a plane with a choice of orthonormal
			basis.
			The solid and dashed arrows show two such bases in an
			oblique view of the same plane.}
		\Description{%
			On the left, a vector e projects orthogonally onto a singular line.
			A dashed perpendicular segment and a right-angle marker
			identify the projection, shown as a green arrow.
			On the right, a shaded plane contains a solid green basis and a
			dashed alternative basis obtained by rotation within the plane.
			Both bases are orthonormal in the plane before projection
			onto the page.}
		\label{fig:projectors}
	\end{figure}
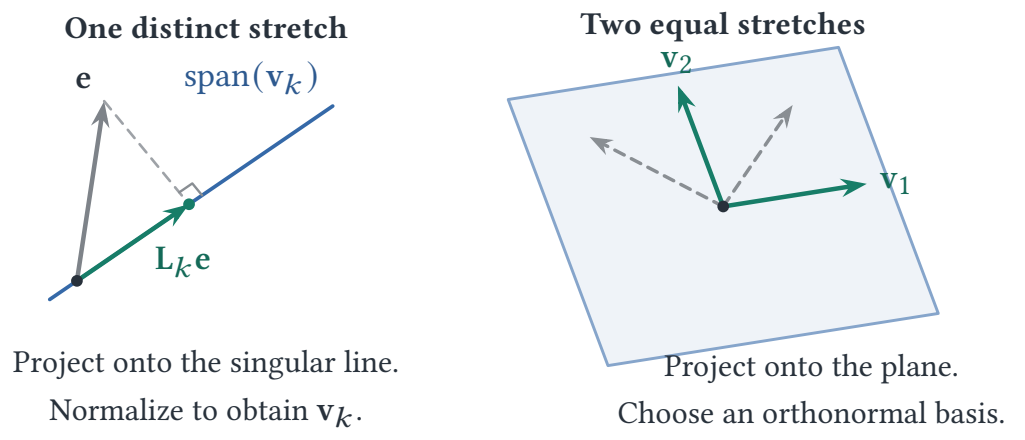

This is the Lagrange basis polynomial of \eqn{lagrange} evaluated at $\mbS$, which is Sylvester's formula for an eigenprojector. The same basis that cancels the difference denominators in the stress therefore also delivers the singular directions. It is one on the $\sigma_k$ direction and zero on the other two, so $\mathbf L_k$ is a rank-one orthogonal projector. Choose a standard basis vector $\mathbf e$ with $\mathbf L_k\mathbf e\ne0$, as on the left of Figure~\ref{fig:projectors}. Then
\begin{equation}\label{eq:svd-vector}
	\mathbf v_k=\frac{\mathbf L_k\mathbf e}{\|\mathbf L_k\mathbf e\|},
	\qquad\mathbf u_k=\mbR\mathbf v_k.
\end{equation}
The three right vectors are orthogonal. Choose their signs so that $\det\mbV=1$. Then $\mbU=\mbR\mbV$ is also proper and
\begin{equation}\label{eq:svd-result}
	\mbF=\mbU\boldsymbol\Sigma\mbV^\top.
\end{equation}
This construction also covers one zero stretch on a regular branch, because it divides by stretch gaps and never by the stretches themselves.

	\paragraph{Equal stretches.}
The individual projectors of an equal pair become ambiguous, as the right of Figure~\ref{fig:projectors} shows, and their sum stays well defined. For $(a,a,b)$ with $a\ne b$ the two eigenspace projectors are
\begin{equation}\label{eq:cluster-projectors}
	\mathbf L_b=\frac{\mbS-a\mbI}{b-a},
	\qquad\mathbf L_a=\mbI-\mathbf L_b.
\end{equation}
The first projects onto the $b$ direction and the second onto the repeated $a$ plane. Orthogonalizing two independent projections of Cartesian basis vectors into the plane gives a valid pair of right singular vectors. A unit vector from $\mathbf L_b$ completes $\mbV$, and $\mbU=\mbR\mbV$ as before.

At a triple coincidence $\mbS=a\mbI$, and any $\mbV\in\SO$ is valid with $\mbU=\mbR\mbV$.

		\subsection{Newton polynomial form}\label{sec:svd-newton}
	
	On a regular branch, where $\alpha\ne0$, the Newton basis recovers the same directions without dividing by stretch gaps. It is the matrix version of Newton's form \eqn{newton}, just as the Lagrange projector is the matrix version of \eqn{lagrange}.
	
	The two matrix factors in Newton's formula are
	\begin{equation}\label{eq:newton-basis-matrices}
		\mathbf N_1=\mbS-\sigma_1\mbI,
		\qquad
		\mathbf N_2=(\mbS-\sigma_1\mbI)(\mbS-\sigma_2\mbI).
	\end{equation}
	
	Their action follows directly from
	$\mbS\mathbf v_k=\sigma_k\mathbf v_k$:
	\begin{equation}\label{eq:newton-basis-action}
		\begin{aligned}
			\mathbf N_1
			&=(\sigma_2-\sigma_1)\mathbf v_2\mathbf v_2^\top
			+(\sigma_3-\sigma_1)\mathbf v_3\mathbf v_3^\top,\\
			\mathbf N_2
			&=(\sigma_3-\sigma_1)(\sigma_3-\sigma_2)
			\mathbf v_3\mathbf v_3^\top.
		\end{aligned}
	\end{equation}
	Thus every nonzero column of $\mathbf N_2$ points along
		$\mathbf v_3$.
		Normalize one such column to obtain this vector.
		Normalization removes the scalar stretch-gap product, and its sign only flips the sign of the vector, which is free anyway.
		The labels are free, so we give the index $3$ to the most isolated stretch, the one with the largest $|\gamma_k|$. This choice makes the scale of $\mathbf N_2$ as large as possible relative to rounding.
	
	Next remove this direction from $\mathbf N_1$:
	\begin{equation}\label{eq:newton-second-direction}
		(\mbI-\mathbf v_3\mathbf v_3^\top)\mathbf N_1
		=(\sigma_2-\sigma_1)\mathbf v_2\mathbf v_2^\top.
	\end{equation}
	Normalize a nonzero column to obtain $\mathbf v_2$.
	Complete the right frame with
	\begin{equation}\label{eq:newton-frame-completion}
		\mathbf v_1=\mathbf v_2\times\mathbf v_3,
		\qquad
		\mbV=[\,\mathbf v_1\ \mathbf v_2\ \mathbf v_3\,].
	\end{equation}
	The cross product ensures $\det\mbV=1$.
	Finally set $\mbU=\mbR\mbV$.
	This gives the signed SVD in \eqn{svd-result}.
	As in the Lagrange construction, it also covers one zero stretch
		on a regular trace branch.
	
	\paragraph{Equal stretches leave a basis choice.}
		If exactly two stretches are equal, order them so that
		$\sigma_1=\sigma_2\ne\sigma_3$. Then
		$\mathbf N_2=(\sigma_3-\sigma_1)^2\mathbf v_3\mathbf v_3^\top$ still recovers the isolated direction,
		and \eqn{newton-second-direction} gives zero,
		which reflects the free basis in the repeated-stretch plane.
		Choose a Cartesian basis vector $\mathbf e$ with a nonzero component orthogonal to $\mathbf v_3$ and set
		\begin{equation}\label{eq:newton-repeated-direction}
			\mathbf v_2
			=
			\frac{(\mbI-\mathbf v_3\mathbf v_3^\top)\mathbf e}
			{\|(\mbI-\mathbf v_3\mathbf v_3^\top)\mathbf e\|},
		\end{equation}
		then complete $\mathbf v_1$ with \eqn{newton-frame-completion}.
		If all three stretches agree, both Newton basis matrices vanish, and $\mbV=\mbI$ with $\mbU=\mbR$ is valid.
	
	\section{Evaluation}\label{sec:svd-evaluation}
We measure the polar formula of \eqn{beta} and the root-gradient, Lagrange and Newton forms of Sections~\ref{sec:svd-root} to~\ref{sec:svd-newton}, in C++ and double precision on a 13th Gen Intel Core i7-1365U CPU. The baseline is the general-purpose Jacobi SVD of Eigen~5.0.1~\citep{Eigen501}, a library of comparable engineering effort. The measurements show what the analytic equations deliver as written, without iterative refinement or SIMD tuning. The quartic $\mathcal{P}$ is solved using the low-rank $LDL^\top$ solver of \citet{Oelen2023LDLT}, which builds on \citet{Strobach2015LDLT}.
	
	\begin{table}[t]
		\centering
		\caption{
				Median time in nanoseconds per $3\times3$ matrix, in double precision over nine timing trials,
				for the polar factors $(\mbR,\mbS)$ and the complete signed SVD $(\mbU,\boldsymbol\Sigma,\mbV)$.
				Parentheses give the speedup over Eigen's Jacobi SVD route for the same output.
			}
			\label{tab:polar-svd-timing}
		\footnotesize
			\setlength{\tabcolsep}{2.4pt}
			\resizebox{\columnwidth}{!}{%
			\begin{tabular}{@{}lrr@{\hspace{9pt}}rrrr@{}}
			\toprule
			& \multicolumn{2}{c}{Polar factors}
			& \multicolumn{4}{c}{Signed SVD} \\
			\cmidrule(lr){2-3}\cmidrule(l){4-7}
			Input family
			& Eigen & Quartic
			& Eigen & Root & Newton & Lagrange \\
			\midrule
			Near identity
			& 617.5 & 420.9 ($1.47\times$)
			& 603.4 & 576.5 ($1.05\times$) & 613.2 ($0.98\times$) & 606.6 ($0.99\times$) \\
			Distinct, positive
			& 717.1 & 459.7 ($1.56\times$)
			& 686.4 & 632.5 ($1.09\times$) & 669.8 ($1.02\times$) & 635.3 ($1.08\times$) \\
			Distinct, inverted
			& 671.2 & 437.6 ($1.53\times$)
			& 665.5 & 588.8 ($1.13\times$) & 629.2 ($1.06\times$) & 615.3 ($1.08\times$) \\
			Clustered
			& 631.8 & 420.8 ($1.50\times$)
			& 621.3 & 550.1 ($1.13\times$) & 613.2 ($1.01\times$) & 591.5 ($1.05\times$) \\
			Near singular
			& 669.9 & 406.8 ($1.65\times$)
			& 647.8 & 550.6 ($1.18\times$) & 595.6 ($1.09\times$) & 572.4 ($1.13\times$) \\
			\bottomrule
		\end{tabular}}
		\end{table}
	
	\paragraph{Cost and computational ordering.} Table~\ref{tab:polar-svd-timing} summarizes the timings. Away from the trace singularity, the quartic route is $1.47$--$1.56\times$ faster than Eigen for polar decomposition. The complete signed SVD is competitive with Eigen, with speedups of $1.05$--$1.13\times$ for the root-gradient form, $0.99$--$1.08\times$ for the Lagrange form and $0.98$--$1.06\times$ for the Newton form. The root-gradient form is fastest because it reuses the products $\mbF\mbC$ and $\operatorname{cof}\mbF$ of the rotation and forms one column of each projector. The quartic route obtains the rotation before the singular vectors, and Eigen works in the opposite order.

	\paragraph{Accuracy.}
	Accuracy is measured by the reconstruction residual $\|\mbU\boldsymbol\Sigma\mbV^\top-\mbF\|_\mathrm{F}/\|\mbF\|_\mathrm{F}$. The clustering tolerances come from the accuracy of the quartic roots. A repeated root is computed less accurately than a simple one, with an error of order $\epsilon^{1/2}$ for a pair and $\epsilon^{1/3}$ for a triple, where $\epsilon=2^{-52}$ is the machine epsilon of double precision. An exact coincidence therefore arrives slightly split. Measured relative to $\max_k|\sigma_k|$, the forms treat three stretches as one when $\sigma_1-\sigma_3\le3\epsilon^{1/3}$, and treat two stretches as one when their gap is at most $4(\epsilon/r)^{1/2}$. Here $r$ is the relative distance from the pair to the third stretch. Over $10^6$ random rotations, the largest observed splits are $1.43\epsilon^{1/3}$ and $1.95(\epsilon/r)^{1/2}$, which stay below these tolerances. The Lagrange and root-gradient forms test the stretch gaps, before any division by $\mathcal P'(\zeta_k)$, and the Newton form tests its matrix products. All three replace a detected cluster by its mean.

Over 200 samples with distinct stretches, positive or inverted, the largest residual is $1.1\times10^{-15}$--$5.0\times10^{-15}$ for the three forms and about $2.3\times10^{-15}$ for Eigen. Near identity, it is $4.1\times10^{-14}$ for the root-gradient and Newton forms, $4.7\times10^{-14}$ for the Lagrange form and $1.6\times10^{-15}$ for Eigen. An exact triple coincidence gives $6.8\times10^{-16}$ for all three forms. For a pair separated by $10^{-7}$, the residual is $2.0\times10^{-9}$ for the root-gradient form, $3.4\times10^{-9}$ for the Lagrange form and $4.7\times10^{-8}$ for the Newton form. Genuine near-triples with gaps of $10^{-5}$ are merged, which gives residuals of $4.2\times10^{-6}$--$8.2\times10^{-6}$. Near rank one, the root-gradient form loses accuracy, because $f-\zeta_k=2(\sigma_i+\sigma_j)$ is small and the isolated companion root lies close to $f$. For signed stretches $(1,10^{-7},5\times10^{-8})$ its residual reaches $1.3\times10^{-2}$, against $3.6\times10^{-8}$ for the Newton form and $6.0\times10^{-8}$ for the Lagrange form.

	\paragraph{Approaching the trace singularity.}
	For $\boldsymbol\Sigma=\operatorname{diag}(2,1,-(1-\delta))$ we have $\alpha=24\delta(1+\delta)$, so $\alpha\to0$ as $\delta\to0$. Over 200 rotation pairs, the largest discrepancy $\|\mbR-\mbR_{\mathrm{Eigen}}\|_{\mathrm{F}}$ is $3.2\times10^{-7}$ at $\delta=10^{-4}$, $2.9\times10^{-3}$ at $\delta=10^{-6}$ and $3.8\times10^{-1}$ at $\delta=10^{-7}$. At $\delta=10^{-6}$ the orthogonality residual $\|\mbR^\top\mbR-\mbI\|_\mathrm{F}$ reaches $5.8\times10^{-3}$. At $\delta=10^{-8}$ each quartic pipeline reports failure on 18 of 200 inputs, because the polar code rejects a nonfinite or vanishing computed $\alpha$ ($|\alpha|\le10^{-300}$), and both pipelines check their factors for finiteness only. These losses occur only as $\alpha$ approaches zero, at the kink of the energy, where the rotation itself is ill-conditioned by \eqn{rotation-norm}. A simulation reaches such inverted configurations typically through solver error. The factors of $\alpha$ are the stretch sums that \citet{Stomakhin2012Supplement} regularize in the stress derivatives. Extending that regularization to the quartic rotation, with a refinement that restores orthogonality, is future work.

	\section{Conclusion}\label{sec:conclusion}
	The characteristic cubic carries the stretch differences of the co-rotational Hessian, and interpolation cancels them in the assembled stress. The trace quartic carries the stretch sums through its derivative at the selected root, and their zeros on the maximum-trace branch are exactly the kinks of the energy, where the nearest proper rotation is non-unique. Averaging the selected rotation with a companion gives each stretch gradient, and the matrix forms of the same interpolation turn these into a signed SVD. The analytic formulas compute polar factors $1.47$--$1.56\times$ faster than Eigen away from the trace singularity, and the root-gradient signed SVD is $1.05$--$1.13\times$ faster.

	In $n$ dimensions the same elimination gives a trace polynomial of degree $2^{n-1}$, which in four dimensions is a quartic in $f^2$. The identity $\mbR=\partial f/\partial\mbF$ and the stretch gradients as rotation averages carry over, and the open question is where the selected root loses simplicity.

		\appendix
	
	\section{The Cubic Within the Quartic}\label{app:quartic-cubic}
The root formulas give an exact connection between the two polynomials,
\begin{equation}\label{eq:quartic-cubic-identity}
\mathcal P(2s-f)=16(s-f)\mathcal C(s).
\end{equation}
The factor $s-f$ belongs to the selected root, and the other three factors recover the signed stretches. Differentiation with respect to $s$ and evaluation at a stretch gives
\begin{equation}\label{eq:quartic-cubic-slopes}
\mathcal P'(\zeta_k)=8(\sigma_k-f)\mathcal C'(\sigma_k).
\end{equation}
This connects \eqn{tau} and \eqn{gamma}.

	\section{Discriminants}\label{app:discriminants}
	A discriminant is the product of all squared root gaps, so it vanishes when any two roots agree.
	
	Let $\Delta_{\mathcal C}=\prod_{i<j}(\sigma_i-\sigma_j)^2$. The eigenvalues of $\mbC$ are $\lambda_k=\sigma_k^2$. Their discriminant is
	\begin{equation}\label{eq:discriminants}
		\boxed{\Delta_\lambda
			=\Delta_{\mathcal C}\left(\frac\alpha8\right)^2
			=\frac{\Delta_{\mathcal P}}{2^{12}}.}
	\end{equation}
	The first equality follows by factoring $\lambda_i-\lambda_j=(\sigma_i-\sigma_j)(\sigma_i+\sigma_j)$. For the second equality use \eqn{roots}. The six quartic gaps are twice the three stretch differences and twice the three stretch sums. Squaring their product supplies the factor $2^{12}$. The quartic discriminant therefore vanishes at equal stretches and at zero sums alike, while $\alpha$ isolates the sums of the selected root.
	
	\bibliographystyle{ACM-Reference-Format}
	\bibliography{corotational-algebra}


\begin{thebibliography}{42}


\ifx \showCODEN    \undefined \def \showCODEN     #1{\unskip}     \fi
\ifx \showDOI      \undefined \def \showDOI       #1{#1}\fi
\ifx \showISBNx    \undefined \def \showISBNx     #1{\unskip}     \fi
\ifx \showISBNxiii \undefined \def \showISBNxiii  #1{\unskip}     \fi
\ifx \showISSN     \undefined \def \showISSN      #1{\unskip}     \fi
\ifx \showLCCN     \undefined \def \showLCCN      #1{\unskip}     \fi
\ifx \shownote     \undefined \def \shownote      #1{#1}          \fi
\ifx \showarticletitle \undefined \def \showarticletitle #1{#1}   \fi
\ifx \showURL      \undefined \def \showURL       {\relax}        \fi
\providecommand\bibfield[2]{#2}
\providecommand\bibinfo[2]{#2}
\providecommand\natexlab[1]{#1}
\providecommand\showeprint[2][]{arXiv:#2}

\bibitem[Alexa et~al\mbox{.}(2023)]%
        {10.1145/3596711.3596731}
\bibfield{author}{\bibinfo{person}{Marc Alexa}, \bibinfo{person}{Daniel
  Cohen-Or}, {and} \bibinfo{person}{David Levin}.}
  \bibinfo{year}{2023}\natexlab{}.
\newblock \bibinfo{booktitle}{\emph{As-Rigid-As-Possible Shape Interpolation}
  (\bibinfo{edition}{1} ed.)}.
\newblock \bibinfo{publisher}{Association for Computing Machinery},
  \bibinfo{address}{New York, NY, USA}.
\newblock
\showISBNx{9798400708978}


\bibitem[Ball(1984)]%
        {Ball1984}
\bibfield{author}{\bibinfo{person}{J.~M. Ball}.}
  \bibinfo{year}{1984}\natexlab{}.
\newblock \showarticletitle{Differentiability Properties of Symmetric and
  Isotropic Functions}.
\newblock \bibinfo{journal}{\emph{Duke Mathematical Journal}}
  \bibinfo{volume}{51}, \bibinfo{number}{3} (\bibinfo{year}{1984}),
  \bibinfo{pages}{699--728}.
\newblock


\bibitem[Chadwick and Ogden(1971)]%
        {Chadwick1971}
\bibfield{author}{\bibinfo{person}{Peter Chadwick} {and}
  \bibinfo{person}{Raymond~W. Ogden}.} \bibinfo{year}{1971}\natexlab{}.
\newblock \showarticletitle{On the Definition of Elastic Moduli}.
\newblock \bibinfo{journal}{\emph{Archive for Rational Mechanics and Analysis}}
  \bibinfo{volume}{44}, \bibinfo{number}{1} (\bibinfo{year}{1971}),
  \bibinfo{pages}{41--53}.
\newblock


\bibitem[Chao et~al\mbox{.}(2010)]%
        {Chao2010}
\bibfield{author}{\bibinfo{person}{Isaac Chao}, \bibinfo{person}{Ulrich
  Pinkall}, \bibinfo{person}{Patrick Sanan}, {and} \bibinfo{person}{Peter
  Schr{\"o}der}.} \bibinfo{year}{2010}\natexlab{}.
\newblock \showarticletitle{A Simple Geometric Model for Elastic Deformations}.
\newblock \bibinfo{journal}{\emph{ACM Transactions on Graphics}}
  \bibinfo{volume}{29}, \bibinfo{number}{4}, Article \bibinfo{articleno}{38}
  (\bibinfo{year}{2010}), \bibinfo{numpages}{6}~pages.
\newblock


\bibitem[Connolly et~al\mbox{.}(2019)]%
        {Connolly2019}
\bibfield{author}{\bibinfo{person}{Stephen~J. Connolly},
  \bibinfo{person}{Donald Mackenzie}, {and} \bibinfo{person}{Yevgen Gorash}.}
  \bibinfo{year}{2019}\natexlab{}.
\newblock \showarticletitle{Isotropic Hyperelasticity in Principal Stretches:
  Explicit Elasticity Tensors and Numerical Implementation}.
\newblock \bibinfo{journal}{\emph{Computational Mechanics}}
  \bibinfo{volume}{64}, \bibinfo{number}{5} (\bibinfo{year}{2019}),
  \bibinfo{pages}{1273--1288}.
\newblock


\bibitem[Danskin(1967)]%
        {Danskin1967}
\bibfield{author}{\bibinfo{person}{John~M. Danskin}.}
  \bibinfo{year}{1967}\natexlab{}.
\newblock \bibinfo{booktitle}{\emph{The Theory of Max-Min and its Application
  to Weapons Allocation Problems}}. \bibinfo{series}{Econometrics and
  Operations Research}, Vol.~\bibinfo{volume}{5}.
\newblock \bibinfo{publisher}{Springer}, \bibinfo{address}{Berlin, Heidelberg}.
\newblock


\bibitem[Franca(1989)]%
        {Franca1989}
\bibfield{author}{\bibinfo{person}{Leopoldo~P. Franca}.}
  \bibinfo{year}{1989}\natexlab{}.
\newblock \showarticletitle{An Algorithm to Compute the Square Root of a
  $3\times3$ Positive Definite Matrix}.
\newblock \bibinfo{journal}{\emph{Computers \& Mathematics with Applications}}
  \bibinfo{volume}{18}, \bibinfo{number}{5} (\bibinfo{year}{1989}),
  \bibinfo{pages}{459--466}.
\newblock


\bibitem[Gawlik and Leok(2017)]%
        {doi:10.1137/16M108971X}
\bibfield{author}{\bibinfo{person}{Evan~S. Gawlik} {and}
  \bibinfo{person}{Melvin Leok}.} \bibinfo{year}{2017}\natexlab{}.
\newblock \showarticletitle{Iterative Computation of the Fréchet Derivative of
  the Polar Decomposition}.
\newblock \bibinfo{journal}{\emph{SIAM J. Matrix Anal. Appl.}}
  \bibinfo{volume}{38}, \bibinfo{number}{4} (\bibinfo{year}{2017}),
  \bibinfo{pages}{1354--1379}.
\newblock


\bibitem[Guennebaud et~al\mbox{.}(2025)]%
        {Eigen501}
\bibfield{author}{\bibinfo{person}{Ga{\"e}l Guennebaud},
  \bibinfo{person}{Beno{\^i}t Jacob}, {et~al\mbox{.}}}
  \bibinfo{year}{2025}\natexlab{}.
\newblock \bibinfo{title}{{Eigen}: A {C++} Template Library for Linear
  Algebra}.
\newblock
\newblock
\newblock
\shownote{Version 5.0.1}.


\bibitem[Higham(2008)]%
        {Higham2008}
\bibfield{author}{\bibinfo{person}{Nicholas~J. Higham}.}
  \bibinfo{year}{2008}\natexlab{}.
\newblock \bibinfo{booktitle}{\emph{Functions of Matrices: Theory and
  Computation}}.
\newblock \bibinfo{publisher}{Society for Industrial and Applied Mathematics},
  \bibinfo{address}{Philadelphia, PA}.
\newblock


\bibitem[Higham and Noferini(2016)]%
        {HighamNoferini2016}
\bibfield{author}{\bibinfo{person}{Nicholas~J. Higham} {and}
  \bibinfo{person}{Vanni Noferini}.} \bibinfo{year}{2016}\natexlab{}.
\newblock \showarticletitle{An Algorithm to Compute the Polar Decomposition of
  a $3\times3$ Matrix}.
\newblock \bibinfo{journal}{\emph{Numerical Algorithms}} \bibinfo{volume}{73},
  \bibinfo{number}{2} (\bibinfo{year}{2016}), \bibinfo{pages}{349--369}.
\newblock


\bibitem[Hoger and Carlson(1984)]%
        {HogerCarlson1984}
\bibfield{author}{\bibinfo{person}{Anne Hoger} {and} \bibinfo{person}{Donald~E.
  Carlson}.} \bibinfo{year}{1984}\natexlab{}.
\newblock \showarticletitle{Determination of the Stretch and Rotation in the
  Polar Decomposition of the Deformation Gradient}.
\newblock \bibinfo{journal}{\emph{Quart. Appl. Math.}} \bibinfo{volume}{42},
  \bibinfo{number}{1} (\bibinfo{year}{1984}), \bibinfo{pages}{113--117}.
\newblock


\bibitem[Horn(1987)]%
        {Horn1987}
\bibfield{author}{\bibinfo{person}{Berthold K.~P. Horn}.}
  \bibinfo{year}{1987}\natexlab{}.
\newblock \showarticletitle{Closed-Form Solution of Absolute Orientation Using
  Unit Quaternions}.
\newblock \bibinfo{journal}{\emph{Journal of the Optical Society of America A}}
  \bibinfo{volume}{4}, \bibinfo{number}{4} (\bibinfo{year}{1987}),
  \bibinfo{pages}{629--642}.
\newblock


\bibitem[Irving et~al\mbox{.}(2004)]%
        {10.1145/1028523.1028541}
\bibfield{author}{\bibinfo{person}{G. Irving}, \bibinfo{person}{J. Teran},
  {and} \bibinfo{person}{R. Fedkiw}.} \bibinfo{year}{2004}\natexlab{}.
\newblock \showarticletitle{Invertible finite elements for robust simulation of
  large deformation}. In \bibinfo{booktitle}{\emph{Proceedings of the 2004 ACM
  SIGGRAPH/Eurographics Symposium on Computer Animation}} (Grenoble, France)
  \emph{(\bibinfo{series}{SCA '04})}. \bibinfo{publisher}{Eurographics
  Association}, \bibinfo{address}{Goslar, DEU}, \bibinfo{pages}{131–140}.
\newblock
\showISBNx{3905673142}


\bibitem[Jeremi{\'c} and Cheng(2005)]%
        {JeremicCheng2005}
\bibfield{author}{\bibinfo{person}{Boris Jeremi{\'c}} {and}
  \bibinfo{person}{Zhao Cheng}.} \bibinfo{year}{2005}\natexlab{}.
\newblock \showarticletitle{Significance of Equal Principal Stretches in
  Computational Hyperelasticity}.
\newblock \bibinfo{journal}{\emph{Communications in Numerical Methods in
  Engineering}} \bibinfo{volume}{21}, \bibinfo{number}{9}
  (\bibinfo{year}{2005}), \bibinfo{pages}{477--486}.
\newblock


\bibitem[Kim and Eberle(2022)]%
        {KimEberle2022}
\bibfield{author}{\bibinfo{person}{Theodore Kim} {and} \bibinfo{person}{David
  Eberle}.} \bibinfo{year}{2022}\natexlab{}.
\newblock \showarticletitle{Dynamic Deformables: Implementation and Production
  Practicalities (Now with Code!)}. In \bibinfo{booktitle}{\emph{ACM SIGGRAPH
  2022 Courses}} \emph{(\bibinfo{series}{SIGGRAPH '22})}.
  \bibinfo{publisher}{Association for Computing Machinery}, Article
  \bibinfo{articleno}{7}, \bibinfo{numpages}{259}~pages.
\newblock


\bibitem[Kopp(2008)]%
        {Kopp2008Hermitian3x3}
\bibfield{author}{\bibinfo{person}{Joachim Kopp}.}
  \bibinfo{year}{2008}\natexlab{}.
\newblock \showarticletitle{Efficient Numerical Diagonalization of Hermitian $3
  \times 3$ Matrices}.
\newblock \bibinfo{journal}{\emph{International Journal of Modern Physics C}}
  \bibinfo{volume}{19}, \bibinfo{number}{3} (\bibinfo{year}{2008}),
  \bibinfo{pages}{523--548}.
\newblock


\bibitem[Lin et~al\mbox{.}(2022)]%
        {Lin2022}
\bibfield{author}{\bibinfo{person}{Huancheng Lin}, \bibinfo{person}{Floyd~M.
  Chitalu}, {and} \bibinfo{person}{Taku Komura}.}
  \bibinfo{year}{2022}\natexlab{}.
\newblock \showarticletitle{Isotropic {ARAP} Energy Using {Cauchy--Green}
  Invariants}.
\newblock \bibinfo{journal}{\emph{ACM Transactions on Graphics}}
  \bibinfo{volume}{41}, \bibinfo{number}{6}, Article \bibinfo{articleno}{275}
  (\bibinfo{year}{2022}), \bibinfo{numpages}{14}~pages.
\newblock


\bibitem[Lin et~al\mbox{.}(2024)]%
        {10.1145/3666086}
\bibfield{author}{\bibinfo{person}{Huancheng Lin}, \bibinfo{person}{Floyd~M.
  Chitalu}, {and} \bibinfo{person}{Taku Komura}.}
  \bibinfo{year}{2024}\natexlab{}.
\newblock \showarticletitle{Analytic rotation-invariant modelling of
  anisotropic finite elements}.
\newblock \bibinfo{journal}{\emph{ACM Trans. Graph.}} \bibinfo{volume}{43},
  \bibinfo{number}{5}, Article \bibinfo{articleno}{157} (\bibinfo{date}{Aug.}
  \bibinfo{year}{2024}), \bibinfo{numpages}{20}~pages.
\newblock
\showISSN{0730-0301}


\bibitem[Mackey(1995)]%
        {mackey1995hamilton}
\bibfield{author}{\bibinfo{person}{Niloufer Mackey}.}
  \bibinfo{year}{1995}\natexlab{}.
\newblock \showarticletitle{Hamilton and Jacobi Meet Again: Quaternions and the
  Eigenvalue Problem}.
\newblock \bibinfo{journal}{\emph{SIAM J. Matrix Anal. Appl.}}
  \bibinfo{volume}{16}, \bibinfo{number}{2} (\bibinfo{year}{1995}),
  \bibinfo{pages}{421--435}.
\newblock


\bibitem[Markley(1993)]%
        {Markley1993FOAM}
\bibfield{author}{\bibinfo{person}{F.~Landis Markley}.}
  \bibinfo{year}{1993}\natexlab{}.
\newblock \showarticletitle{Attitude Determination Using Vector Observations: A
  Fast Optimal Matrix Algorithm}.
\newblock \bibinfo{journal}{\emph{The Journal of the Astronautical Sciences}}
  \bibinfo{volume}{41}, \bibinfo{number}{2} (\bibinfo{year}{1993}),
  \bibinfo{pages}{261--280}.
\newblock


\bibitem[McAdams et~al\mbox{.}(2011)]%
        {10.1145/2010324.1964932}
\bibfield{author}{\bibinfo{person}{Aleka McAdams}, \bibinfo{person}{Yongning
  Zhu}, \bibinfo{person}{Andrew Selle}, \bibinfo{person}{Mark Empey},
  \bibinfo{person}{Rasmus Tamstorf}, \bibinfo{person}{Joseph Teran}, {and}
  \bibinfo{person}{Eftychios Sifakis}.} \bibinfo{year}{2011}\natexlab{}.
\newblock \showarticletitle{Efficient elasticity for character skinning with
  contact and collisions}.
\newblock \bibinfo{journal}{\emph{ACM Trans. Graph.}} \bibinfo{volume}{30},
  \bibinfo{number}{4}, Article \bibinfo{articleno}{37} (\bibinfo{date}{July}
  \bibinfo{year}{2011}), \bibinfo{numpages}{12}~pages.
\newblock
\showISSN{0730-0301}


\bibitem[Miehe(1993)]%
        {Miehe1993}
\bibfield{author}{\bibinfo{person}{Christian Miehe}.}
  \bibinfo{year}{1993}\natexlab{}.
\newblock \showarticletitle{Computation of Isotropic Tensor Functions}.
\newblock \bibinfo{journal}{\emph{Communications in Numerical Methods in
  Engineering}} \bibinfo{volume}{9}, \bibinfo{number}{11}
  (\bibinfo{year}{1993}), \bibinfo{pages}{889--896}.
\newblock


\bibitem[Miehe(1994)]%
        {Miehe1994}
\bibfield{author}{\bibinfo{person}{Christian Miehe}.}
  \bibinfo{year}{1994}\natexlab{}.
\newblock \showarticletitle{Aspects of the Formulation and Finite Element
  Implementation of Large Strain Isotropic Elasticity}.
\newblock \bibinfo{journal}{\emph{Internat. J. Numer. Methods Engrg.}}
  \bibinfo{volume}{37}, \bibinfo{number}{12} (\bibinfo{year}{1994}),
  \bibinfo{pages}{1981--2004}.
\newblock


\bibitem[M{\"u}ller et~al\mbox{.}(2016)]%
        {10.1145/2994258.2994269}
\bibfield{author}{\bibinfo{person}{Matthias M{\"u}ller}, \bibinfo{person}{Jan
  Bender}, \bibinfo{person}{Nuttapong Chentanez}, {and} \bibinfo{person}{Miles
  Macklin}.} \bibinfo{year}{2016}\natexlab{}.
\newblock \showarticletitle{A robust method to extract the rotational part of
  deformations}. In \bibinfo{booktitle}{\emph{Proceedings of the 9th
  International Conference on Motion in Games}} (Burlingame, California)
  \emph{(\bibinfo{series}{MIG '16})}. \bibinfo{publisher}{Association for
  Computing Machinery}, \bibinfo{address}{New York, NY, USA},
  \bibinfo{pages}{55–60}.
\newblock
\showISBNx{9781450345927}


\bibitem[Myronenko and Song(2009)]%
        {MyronenkoSong2009}
\bibfield{author}{\bibinfo{person}{Andriy Myronenko} {and}
  \bibinfo{person}{Xubo Song}.} \bibinfo{year}{2009}\natexlab{}.
\newblock \bibinfo{title}{On the Closed-Form Solution of the Rotation Matrix
  Arising in Computer Vision Problems}.
\newblock \bibinfo{howpublished}{arXiv:0904.1613}.
\newblock
\showeprint[arxiv]{0904.1613}~[cs.CV]


\bibitem[Neumaier(2001)]%
        {Neumaier_2001}
\bibfield{author}{\bibinfo{person}{Arnold Neumaier}.}
  \bibinfo{year}{2001}\natexlab{}.
\newblock \bibinfo{booktitle}{\emph{Interpolation and Numerical
  Differentiation}}.
\newblock \bibinfo{publisher}{Cambridge University Press},
  \bibinfo{pages}{130–178}.
\newblock


\bibitem[Oelen(2023)]%
        {Oelen2023LDLT}
\bibfield{author}{\bibinfo{person}{Wilco Oelen}.}
  \bibinfo{year}{2023}\natexlab{}.
\newblock \bibinfo{title}{Fast, robust, and accurate low-rank {LDLT} quartic
  equation solver}.  (\bibinfo{year}{2023}).
\newblock
\newblock
\shownote{Manuscript, May 2023}.


\bibitem[Schroeder(2022)]%
        {10.1145/3532720.3535643}
\bibfield{author}{\bibinfo{person}{Craig Schroeder}.}
  \bibinfo{year}{2022}\natexlab{}.
\newblock \showarticletitle{Practical course on computing derivatives in code}.
  In \bibinfo{booktitle}{\emph{ACM SIGGRAPH 2022 Courses}} (Vancouver, British
  Columbia, Canada) \emph{(\bibinfo{series}{SIGGRAPH '22})}.
  \bibinfo{publisher}{Association for Computing Machinery},
  \bibinfo{address}{New York, NY, USA}, Article \bibinfo{articleno}{12},
  \bibinfo{numpages}{28}~pages.
\newblock
\showISBNx{9781450393621}


\bibitem[Shuster and Oh(1981)]%
        {shuster1981three}
\bibfield{author}{\bibinfo{person}{Malcolm~D. Shuster} {and}
  \bibinfo{person}{Seon~Do Oh}.} \bibinfo{year}{1981}\natexlab{}.
\newblock \showarticletitle{Three-Axis Attitude Determination from Vector
  Observations}.
\newblock \bibinfo{journal}{\emph{Journal of Guidance and Control}}
  \bibinfo{volume}{4}, \bibinfo{number}{1} (\bibinfo{date}{Jan.}
  \bibinfo{year}{1981}), \bibinfo{pages}{70--77}.
\newblock


\bibitem[Smith et~al\mbox{.}(2018)]%
        {Smith2018}
\bibfield{author}{\bibinfo{person}{Breannan Smith}, \bibinfo{person}{Fernando
  de Goes}, {and} \bibinfo{person}{Theodore Kim}.}
  \bibinfo{year}{2018}\natexlab{}.
\newblock \showarticletitle{Stable {Neo-Hookean} Flesh Simulation}.
\newblock \bibinfo{journal}{\emph{ACM Transactions on Graphics}}
  \bibinfo{volume}{37}, \bibinfo{number}{2}, Article \bibinfo{articleno}{12}
  (\bibinfo{year}{2018}), \bibinfo{numpages}{15}~pages.
\newblock


\bibitem[Smith et~al\mbox{.}(2019)]%
        {Smith2019}
\bibfield{author}{\bibinfo{person}{Breannan Smith}, \bibinfo{person}{Fernando
  de Goes}, {and} \bibinfo{person}{Theodore Kim}.}
  \bibinfo{year}{2019}\natexlab{}.
\newblock \showarticletitle{Analytic Eigensystems for Isotropic Distortion
  Energies}.
\newblock \bibinfo{journal}{\emph{ACM Transactions on Graphics}}
  \bibinfo{volume}{38}, \bibinfo{number}{1}, Article \bibinfo{articleno}{3}
  (\bibinfo{year}{2019}), \bibinfo{numpages}{15}~pages.
\newblock


\bibitem[Smith(1961)]%
        {10.1145/355578.366316}
\bibfield{author}{\bibinfo{person}{Oliver~K. Smith}.}
  \bibinfo{year}{1961}\natexlab{}.
\newblock \showarticletitle{Eigenvalues of a symmetric 3 \texttimes{} 3
  matrix}.
\newblock \bibinfo{journal}{\emph{Commun. ACM}} \bibinfo{volume}{4},
  \bibinfo{number}{4} (\bibinfo{date}{April} \bibinfo{year}{1961}),
  \bibinfo{pages}{168}.
\newblock
\showISSN{0001-0782}


\bibitem[Sorkine and Alexa(2007)]%
        {10.5555/1281991.1282006}
\bibfield{author}{\bibinfo{person}{Olga Sorkine} {and} \bibinfo{person}{Marc
  Alexa}.} \bibinfo{year}{2007}\natexlab{}.
\newblock \showarticletitle{As-Rigid-as-Possible Surface Modeling}. In
  \bibinfo{booktitle}{\emph{Proceedings of the Fifth Eurographics Symposium on
  Geometry Processing}} (Barcelona, Spain) \emph{(\bibinfo{series}{SGP '07})}.
  \bibinfo{publisher}{Eurographics Association}, \bibinfo{address}{Goslar,
  DEU}, \bibinfo{pages}{109–116}.
\newblock
\showISBNx{9783905673463}


\bibitem[Stewart(1992)]%
        {Stewart1992}
\bibfield{author}{\bibinfo{person}{G.~W. Stewart}.}
  \bibinfo{year}{1992}\natexlab{}.
\newblock \bibinfo{booktitle}{\emph{On the Early History of the Singular Value
  Decomposition}}.
\newblock \bibinfo{type}{{T}echnical {R}eport} CS-TR-2855 / UMIACS-TR-92-31.
  \bibinfo{institution}{University of Maryland}.
\newblock


\bibitem[Stomakhin et~al\mbox{.}(2012a)]%
        {Stomakhin2012}
\bibfield{author}{\bibinfo{person}{Alexey Stomakhin}, \bibinfo{person}{Russell
  Howes}, \bibinfo{person}{Craig Schroeder}, {and} \bibinfo{person}{Joseph~M.
  Teran}.} \bibinfo{year}{2012}\natexlab{a}.
\newblock \showarticletitle{Energetically Consistent Invertible Elasticity}. In
  \bibinfo{booktitle}{\emph{Proceedings of the ACM SIGGRAPH/Eurographics
  Symposium on Computer Animation}} \emph{(\bibinfo{series}{SCA '12})}.
  \bibinfo{publisher}{Eurographics Association}, \bibinfo{pages}{25--32}.
\newblock


\bibitem[Stomakhin et~al\mbox{.}(2012b)]%
        {Stomakhin2012Supplementary}
\bibfield{author}{\bibinfo{person}{Alexey Stomakhin}, \bibinfo{person}{Russell
  Howes}, \bibinfo{person}{Craig Schroeder}, {and} \bibinfo{person}{Joseph~M.
  Teran}.} \bibinfo{year}{2012}\natexlab{b}.
\newblock \bibinfo{booktitle}{\emph{Energetically Consistent Invertible
  Elasticity: Supplementary Technical Document}}.
\newblock \bibinfo{type}{{T}echnical {R}eport}.
  \bibinfo{institution}{University of California, Los Angeles}.
\newblock
\urldef\tempurl%
\url{https://alexey.stomakhin.com/research/sca2012_tech_report.pdf}
\showURL{%
\tempurl}


\bibitem[Stomakhin et~al\mbox{.}(2012c)]%
        {Stomakhin2012Supplement}
\bibfield{author}{\bibinfo{person}{Alexey Stomakhin}, \bibinfo{person}{Russell
  Howes}, \bibinfo{person}{Craig Schroeder}, {and} \bibinfo{person}{Joseph~M.
  Teran}.} \bibinfo{year}{2012}\natexlab{c}.
\newblock \bibinfo{title}{Supplementary Technical Document for ``Energetically
  Consistent Invertible Elasticity''}.
\newblock \bibinfo{howpublished}{Technical supplement}.
\newblock


\bibitem[Strobach(2015)]%
        {Strobach2015LDLT}
\bibfield{author}{\bibinfo{person}{Peter Strobach}.}
  \bibinfo{year}{2015}\natexlab{}.
\newblock \bibinfo{booktitle}{\emph{The Low-Rank {LDLT} Quartic Solver}}.
\newblock \bibinfo{type}{Internal Technical Report}.
  \bibinfo{institution}{AST-Consulting Inc., R{\"o}hrnbach, Germany}.
\newblock


\bibitem[Szabó(2026)]%
        {SZABO2026113971}
\bibfield{author}{\bibinfo{person}{László Szabó}.}
  \bibinfo{year}{2026}\natexlab{}.
\newblock \showarticletitle{On the direct determination of the right stretch
  tensor and the admissible region of its principal invariants}.
\newblock \bibinfo{journal}{\emph{International Journal of Solids and
  Structures}}  \bibinfo{volume}{333} (\bibinfo{year}{2026}),
  \bibinfo{pages}{113971}.
\newblock
\showISSN{0020-7683}


\bibitem[Teran et~al\mbox{.}(2003)]%
        {846276.846285}
\bibfield{author}{\bibinfo{person}{J. Teran}, \bibinfo{person}{S. Blemker},
  \bibinfo{person}{V.~Ng~Thow Hing}, {and} \bibinfo{person}{R. Fedkiw}.}
  \bibinfo{year}{2003}\natexlab{}.
\newblock \showarticletitle{Finite volume methods for the simulation of
  skeletal muscle}. In \bibinfo{booktitle}{\emph{Proceedings of the 2003 ACM
  SIGGRAPH/Eurographics Symposium on Computer Animation}} (San Diego,
  California) \emph{(\bibinfo{series}{SCA '03})}.
  \bibinfo{publisher}{Eurographics Association}, \bibinfo{address}{Goslar,
  DEU}, \bibinfo{pages}{68–74}.
\newblock
\showISBNx{1581136595}


\bibitem[Teran et~al\mbox{.}(2005)]%
        {Teran2005}
\bibfield{author}{\bibinfo{person}{Joseph Teran}, \bibinfo{person}{Eftychios
  Sifakis}, \bibinfo{person}{Geoffrey Irving}, {and} \bibinfo{person}{Ronald
  Fedkiw}.} \bibinfo{year}{2005}\natexlab{}.
\newblock \showarticletitle{Robust Quasistatic Finite Elements and Flesh
  Simulation}. In \bibinfo{booktitle}{\emph{Proceedings of the 2005 ACM
  SIGGRAPH/Eurographics Symposium on Computer Animation}}
  \emph{(\bibinfo{series}{SCA '05})}. \bibinfo{publisher}{Association for
  Computing Machinery}, \bibinfo{pages}{181--190}.
\newblock


\end{thebibliography}
\end{document}